\documentclass[fleqn,10pt]{wlscirep}
\usepackage[utf8]{inputenc}
\usepackage[T1]{fontenc}

\newcommand{\mnras}{MNRAS}
\newcommand{\nat}{Nature}
\newcommand{\apj}{ApJ}
\newcommand{\apjl}{ApJL}
\newcommand{\aap}{A\&A}
\newcommand{\apjs}{ApJS}
\newcommand{\aj}{AJ}
\newcommand{\prd}{Phys. Rev. D}
\newcommand{\araa}{ARA\&A}
\newcommand{\pasp}{PASP}
\newcommand{\pasj}{PASJ}
\newcommand{\ssr}{Space Sci. Rev.}
\newcommand{\jqsrt}{J. Quant. Spectrosc. Radiat. Transf.}

\newcommand{\JournalTitle}[1]{#1}

\newcommand{\nhi}{N_\text{HI}}
\newcommand{\lal}{Lyman-$\alpha$ }

\title{Nearly pristine gas in a galaxy 1 billion years after the Big Bang}

\author[1,2,$\dagger$,*]{Sarah E. I. Bosman}
\author[2,$\dagger$]{Frederick B. Davies}
\author[3,4]{Louise Welsh}
\author[5]{Emanuele P. Farina}
\author[6]{Prakash Gaikwad}
\author[1]{Benedetta Spina}
\author[2]{Barun Maity}
\author[3]{Ryan Cooke}
\author[1]{Klaudia Protušová}
\author[4,7]{Valentina D'Odorico}

\author[2]{Eduardo Ba\~nados}
\author[8]{James S. Bolton}
\author[9]{Rebecca L. Davies}
\author[10]{Xiaohui Fan}
\author[11]{Martin G. Haehnelt}
\author[12,13]{Joseph F. Hennawi}
\author[9]{Tahlia J. Hill}
\author[1,2]{Seyeon Hwang}
\author[14]{Vid Ir\v{s}i\v{c}}
\author[15]{Laura Keating}
\author[16]{Girish Kulkarni}
\author[17]{Samuel Lai}
\author[9]{Emma V. Ryan-Weber}
\author[1]{Lasse Seyberlich}
\author[18]{Molly Wolfson}

\affil[1]{Institute for Theoretical Physics, University of Heidelberg, Philosophenweg 12, D-69115 Heidelberg, Germany}
\affil[2]{Max-Planck-Institut f\"{u}r Astronomie, K\"{o}nigstuhl 17, D-69117 Heidelberg, Germany}
\affil[3]{Centre for Extragalactic Astronomy, Durham University, South Road, Durham DH1 3LE, UK}
\affil[4]{INAF - Osservatorio Astronomico di Trieste, via G. B. Tiepolo 11, I-34143 Trieste, Italy}
\affil[5]{International Gemini Observatory/NSF NOIRLab, 670 N A'ohoku Place, Hilo, Hawai'i 96720, USA}
\affil[6]{Department of Astronomy, Astrophysics and Space Engineering, Indian Institute of Technology Indore, MP, 453552, India}
\affil[7]{IFPU $-$ Institute for Fundamental Physics of the Universe, via Beirut 2, I-34151 Trieste, Italy}
\affil[8]{School of Physics and Astronomy, University of Nottingham, University Park, Nottingham, NG7 2RD, UK}
\affil[9]{Centre for Astrophysics and Supercomputing, Swinburne University of Technology, John Street, Hawthorn, 3122, VIC, Australia}
\affil[10]{Steward Observatory, University of Arizona, 933 N. Cherry Ave, Tucson, AZ 85721, USA}
\affil[11]{Kavli Institute for Cosmology and Institute of Astronomy, Madingley Road, Cambridge, CB3 0HA, UK}
\affil[12]{Leiden Observatory, Leiden University, P.O. Box 9513, NL-2300 RA Leiden, the Netherlands}
\affil[13]{Department of Physics, University of California, Santa Barbara, CA 93106, USA}
\affil[14]{Centre for Astrophysics Research, Department of Physics, Astronomy and Mathematics, University of Hertfordshire, College Lane, Hatfield, AL10 9AB, UK}
\affil[15]{Institute for Astronomy, University of Edinburgh, Blackford Hill, Edinburgh, EH9 3HJ, UK}
\affil[16]{Tata Institute of Fundamental Research, Homi Bhabha Road, Mumbai 400005, India}
\affil[17]{Space \& Astronomy, Commonwealth Scientific and Industrial Research Organisation (CSIRO), P. O. Box 1130, Bentley, WA 6102, Australia}
\affil[18]{Center for Cosmology and AstroParticle Physics, The Ohio State University, 191 West Woodruff Avenue, Columbus, OH 43210, USA}

\affil[$\dagger$]{Equal Contributors}

\affil[*]{bosman@thphys.uni-heidelberg.de}

\begin{document}

\maketitle

\smallskip
\textbf{\Large{Abstract}}
\smallskip

{
Excitation of carbon and oxygen ions is the primary mechanism by which collapsing gas clouds dissipate heat to form new stars. At early cosmic times, in galaxies with chemically pristine gas, many stars are expected to have masses far larger than the mass of our sun due to the deficit in these elements\cite{Bromm03,Frebel07}.
We searched for ultra-metal-poor galaxies using their spectroscopic absorption signatures in front of bright background quasars, employing the new GHOST high-resolution spectrograph\cite{GHOST1,GHOST2}. 
Here we report the discovery of the first galaxy containing a reservoir of neutral hydrogen gas with metallicity below the ultra-metal-poor threshold: with a measured carbon-to-hydrogen ratio 11,000 times lower than the Sun, it is
the most pristine ever found by a factor of four\cite{Welsh23}.
UMP-DLA-1 is the only known galaxy with measured carbon and oxygen abundances too low to enable efficient formation of stars with solar mass or below. 
The low metallicity alone therefore guarantees a top-heavy stellar population, providing a direct view of the chemical composition imprinted by massive low-metallicity stars that perished long before the present day\cite{Bromm11,Marks12}: an elevated oxygen-to-carbon ratio. 
The abundance pattern of UMP-DLA-1 is likely a close representation of the atomic yields of first- or early second-generation stars.}

\smallskip
\textbf{\Large{Main}}
\smallskip

Following Big Bang nucleosynthesis, the baryons in the Universe were predominantly composed of hydrogen and helium, with only minute amounts of heavier elements. The gas clouds that collapsed into the first generation of stars (so-called Population III or Pop III stars) could rely only on H$_2$ and HD to dissipate potential energy. The resulting characteristic fragmentation scales, thought to be of $\gtrsim$10$M_\odot$, lead to a stellar initial mass function (IMF) tilted towards higher mass stars compared to that observed in the present day. 
While the exact slopes and characteristic masses of the IMF as a function of metallicity are still unknown, it is broadly expected that 
second-generation stars (so-called Pop II stars) will continue to follow a top-heavy 
(``log-flat'') IMF 
even after baryons from which they form are modestly enriched by heavy elements.  
Specifically, the IMF remains top-heavy until sufficient amounts of carbon and oxygen (and to a lesser extent, silicon and iron) have been formed and dispersed by Pop III and early Pop II supernovae (SNe) to enable efficient metal-line cooling in collapsing gas. The degree of enrichment required for metal line cooling to be efficient is very low, roughly [C/H]$\gtrsim-3.5$\cite{Bromm03,Santoro06,Bromm11,Marks12,Klessen23}.  

\smallskip
\textbf{\Large{Finding the most pristine galaxies}}
\smallskip

Enrichment to this threshold occurred so rapidly that all high-redshift galaxies detected in emission, including by the \textit{James Webb Space Telescope}, are already enriched far above their primordial state by the time they are bright enough to be observed\cite{Nakajima25,Cullen25}. 
Absorption-selected systems, on the other hand, can be arbitrarily faint. In particular, the most dense absorbing clouds with hydrogen column densities $N_\text{HI}>10^{20.3}$ cm${}^{-2}$ (damped Lyman-$\alpha$ absorbers; DLAs) 
are thought to reside within at most a few kpc from star-forming regions \cite{Wolfe86,Krogager17}, although some debate on this point exists, especially for high-metallicity systems\cite{Neeleman17,Neeleman25}. 
Since their discovery in 1972\cite{Beaver72}, much effort has been undertaken to identify the most metal-poor DLAs -- thus far, the most pristine systems have metal abundances just under 1/1000 of the abundances measured in the Sun \cite{Cooke11,Welsh23}. 
All DLAs found until now have carbon and oxygen abundances sufficient for efficient gas cooling and formation of stars in a normal IMF. Here, we report the discovery of UMP--DLA--1 at $z\simeq5.14$, the first DLA below this so-called ``transition discriminant''\cite{Frebel07}. Unlike any galaxy detected previously, the enrichment pattern in UMP--DLA--1 is certain $-$ due to the object's low metallicity alone $-$ to have been produced by SNe from a low-metallicity, top-heavy IMF. Its chemical signature reflects either the yields of Pop III SNe, or at the very least ultra-low-metallicity Pop II SNe.

\begin{figure}[t]
\begin{center}
\resizebox{18cm}{!}{\includegraphics[trim={0 0 0 0},clip]{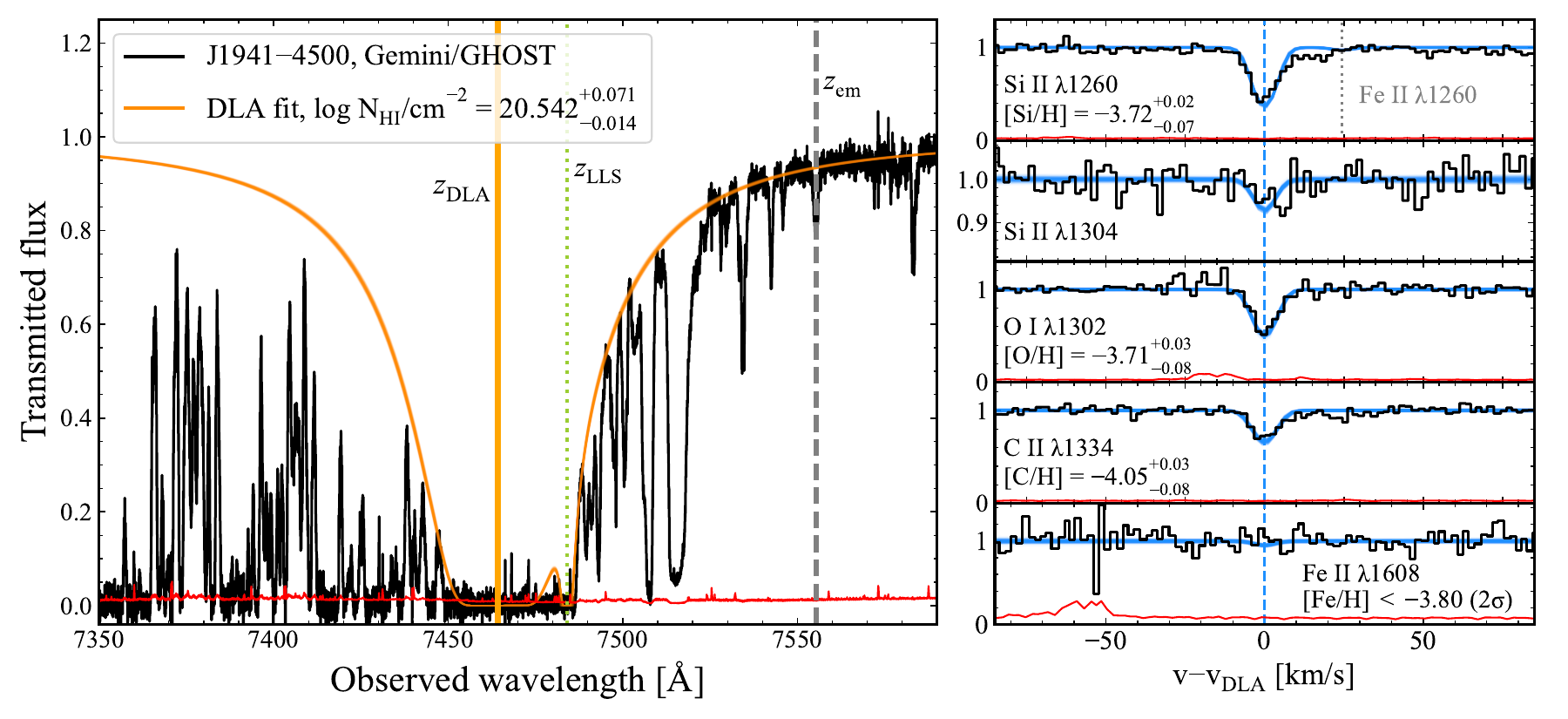}}
\end{center}
\caption{\textbf{Absorption lines in the ultra-metal-poor DLA.} Black and red lines show the measured spectrum and its uncertainty, normalised by the underlying quasar emission continuum, while the coloured curves show 100 draws from the model fit posteriors. Panel (a): absorption profile with the best-fit column density of hydrogen in the DLA, $\log{N_\text{HI}}/$cm$^{-2}=20.542_{-0.014}^{+0.071}$, demonstrating that the system is self-shielded and thus associated with a galaxy. The Ly$\alpha$ redshifts of the quasar ($z_\text{em}$), the DLA ($z_\text{DLA}$) and a LLS in-between ($z_\text{LLS}$) are indicated. Panel (b): detected metal absorption lines and corresponding best-fit metal abundances relative to solar: the carbon, oxygen and silicon abundances are the lowest ever observed by factors $3-4$. The carbon abundance being less than $-$4 establishes the object as the first ``ultra'' metal-poor DLA.
}
\label{fig:spec}
\end{figure}

\begin{figure}[t]
\begin{center}
\resizebox{10cm}{!}{\includegraphics[trim={0 0 0 0},clip]{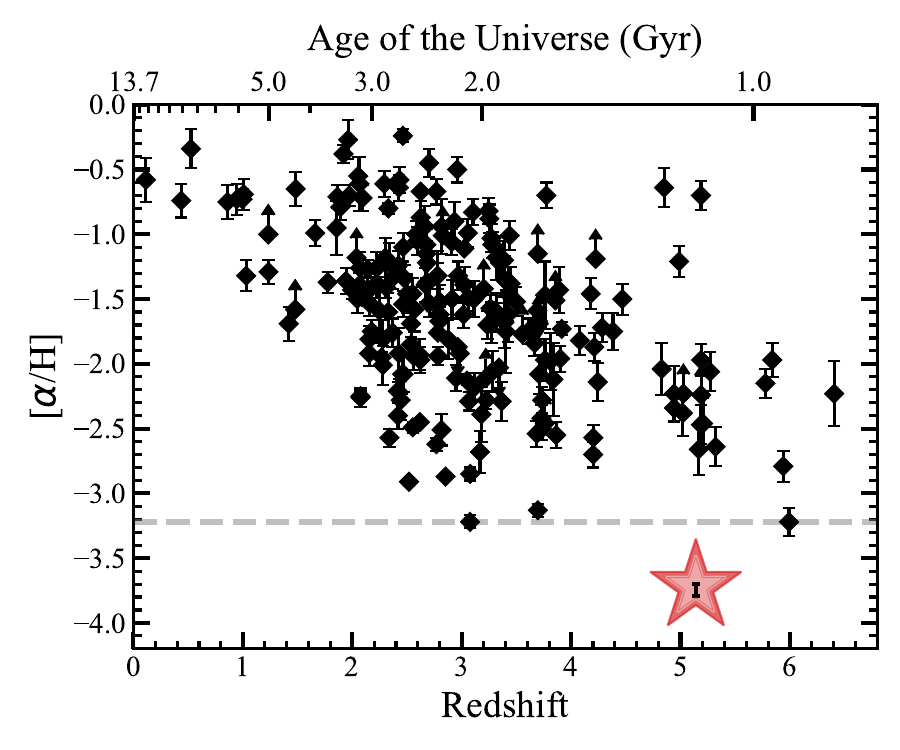}}
\end{center}
\caption{\textbf{Redshift evolution of the $\boldsymbol{\alpha}$-element abundance of DLAs.} UMP--DLA--1 (black point inside the large red star) displays an abundance of $\alpha$ element to hydrogen 5000 times below solar and over three times lower than any other DLA (black points). Its redshift of 5.14 suggests the presence of a metallicity floor which increases with cosmic time. The dashed gray line indicates the previously known most metal-poor DLA, [$\alpha$/H]$=-$3.22.
}
\label{fig:redshift}
\end{figure}

UMP--DLA--1 was discovered serendipitously in front of the background quasar J1941$-$4500 at $z_{\rm em}=5.215\pm0.004$\cite{Schindler20}. The quasar was included in our ongoing large observational program, ``GHOSTLy'', as one of the 30 brightest quasars at $4.5<z<5.5$ without existing deep high-resolution spectroscopy ($R>30,\!000$). We employed the GHOST instrument commissioned in 2024 at the Gemini-South Observatory facility, which is the most sensitive high-resolution spectrograph built to date\cite{GHOST1,GHOST2}. More details of the observations and data reductions are given in the Methods.

UMP--DLA--1's broad damped Lyman-$\alpha$ absorption profile was detected at $z\simeq5.14$, with corresponding low-ionisation absorption lines of Si~{\small{II}} $\lambda1260$\AA, O~{\small{I}} $\lambda 1302$\AA, and C~{\small{II}} $\lambda 1334$ \AA \ detected at SNR$>10$, while Si~{\small{II}} $\lambda\lambda 1304, 1526$\AA, Fe~{\small{II}} $\lambda\lambda 1260, 1608$\AA, and high-ionisation metal absorption lines are not detected. 
As shown in Figure 1, the red side of the damped Lyman-$\alpha$ profile overlaps with the quasar continuum which is unaffected by the Lyman-$\alpha$ forest and therefore with the broad quasar Lyman-$\alpha$ emission line. 
Since the separation between the DLA and the quasar is $v=3660\pm250$ km s${}^{-1}$, the system is not a proximate DLA, which are defined as lying less than $3000$ km s${}^{-1}$ from a quasar\cite{Wolfe86,Ellison02}. 
A Lyman-limit system (LLS) with $N_{\text{HI}}\approx10^{18.5}$\,cm$^{-2}$ is located at $z=5.156$, in-between the DLA and the quasar ($\Delta v = 780$km s${}^{-1}$ from the DLA). Only high-ionisation metal absorption lines of C~{\small{IV}} $\lambda\lambda 1548,1550$\AA \ and Si~{\small{IV}} $\lambda\lambda 1393,1402$\AA \ are strongly detected in the LLS, and the system is consistent with the observed truncation of the quasar spectrum at the Lyman limit. We investigate whether the presence of the LLS may be shielding UMP$-$DLA$-$1 from the quasar's radiation and affecting its ionisation state, but conclude that this is very unlikely unless the DLA's physical density is unexpectedly low (see Methods). Since the LLS's \lal absorption nearly overlaps with the DLA's, we model their column densities together.  

We first measured a precise absorption redshift, as well as Si~{\small II}, O~{\small I}, and C~{\small II} column densities, from the metal absorption features. The lack of strong Si~{\small{II}} $\lambda\lambda 1304, 1526$\AA \ confirms that metal absorption is not saturated, and we therefore fit the metal absorption lines with a single physical component while accounting for both turbulent and thermal broadening separately. We then proceed to fit for the H~{\small I} column density of UMP--DLA--1, fixing the absorption redshift to the best-fit value of $z=5.139987\pm0.000002$. We model the intrinsic quasar emission as a local power-law accompanied by a variable number of Gaussians to parametrise the broad Lyman-$\alpha$ emission line, as well as one additional Gaussian at longer wavelengths to account for any contribution from the broad N~{\small{V}} emission line. The properties of UMP--DLA--1, the high-ionisation LLS, and the quasar continuum parameters are fit concurrently, and we marginalise over both the number of Gaussians and the choice of pixels included in the fit, as described in the Methods and further online material. The resulting hydrogen column density, $\log N_\text{HI}/\text{cm}^{-2} = 20.542_{-0.014}^{+0.071}$, saturation in all transitions of the Lyman-series, and the presence of O~{\small{I}} which is in charge-exchange equilibrium with neutral hydrogen while no high-ionisation absorbers are present, together confirm that the DLA system is self-shielded, the ionisation correction is negligibly small, and the gas is predominantly in the neutral phase. 
In this neutral phase, the low-ionization species we detect account for the vast majority of the total element abundances. 
We obtain metal abundances relative to solar of [Si/H]$=-3.72_{-0.07}^{+0.02}$, [O/H]$=-3.71_{-0.08}^{+0.03}$, and [C/H]$=-4.05_{-0.08}^{+0.03}$. 
The total metallicity of the system, assuming that non-detected elements trace the relative abundance of silicon, is $Z/Z_\odot = -3.76_{-0.08}^{+0.03}$, corresponding to a metal mass fraction of $3.16 \times10^{-6}$. 
This low level of carbon enrichment, less than 1/11,000${}^\text{th}$ of the carbon abundance found in the Sun, makes UMP--DLA--1 the first ultra-metal-poor (UMP) DLA $-$ a threshold defined at [X/H]$<-4$\cite{BeersChristlieb05}.

We obtain an upper limit on the gas temperature $T<4337\ (10,\!368)$\,K at $1\sigma (2\sigma)$ respectively, indicating that the line broadening of the heavy elements is dominated by turbulent motion with a Doppler parameter $b=4.60_{-0.35}^{+0.25}$\,km s${}^{-1}$. The thermal broadening limit is consistent with the saturated troughs observed in the higher order Lyman series transitions of UMP--DLA--1, which independently favour $T<12000$\,K (see Methods). 
While the turbulent broadening is typical for metal-poor DLAs at $2<z<3.5$, the gas temperature shows a slight preference for colder temperatures than the typical $T\sim9600\pm2500$\,K\cite{Cooke15}.

To put UMP--DLA--1 into context, we 
compare its abundances to previous measurements in metal-poor DLAs from the literature. 
So-called ``alpha'' elements are the most abundant product of fusion via chain capture of helium, and are best traced by oxygen, sulfur, and silicon. The most alpha-poor DLAs found until now were towards J0903+2628 at $z=3.08$\cite{Welsh23} and towards PSO~J007+04 at $z=5.99$\cite{Sodini24}, both with [$\alpha$/H]$=-3.22$ as shown in Figure 2. UMP--DLA--1 is therefore over three times more metal-poor in alpha elements, and four times more metal-poor in carbon, than any DLA previously known. 
Non-detection of Fe~{\small{II}} absorption in UMP--DLA--1 yields a limit [Fe/H]$<\!-3.80$ at $2\sigma$. UMP--DLA--1 is therefore also the most Fe-poor DLA found to date; based on the enrichment patterns observed in extremely metal-poor (EMP) DLAs and their scatter, its iron abundance is expected to be [Fe/H]$\simeq\!-$4.3$\pm$0.2.

\smallskip
\textbf{\Large{A direct view of nearly metal-free gas}}
\smallskip

The low level of enrichment measured in UMP--DLA--1 provides a strong constraint on the ``metallicity floor'' of the intergalactic medium from which its gas originated, which can be no higher than [C/H]$\simeq\!-4$. 
Although metal enrichment of the intergalactic medium at $z=5$ is thought to be inhomogeneous\cite{Booth12,Kirihara20}, this local limit implies a global limit since we found at least one region in the Universe in which the metal enrichment is $1/11,000^{\text{th}}$ of the solar value.

\begin{figure}[t]
\begin{center}
\resizebox{10cm}{!}{\includegraphics[trim={0 0 0 0},clip]{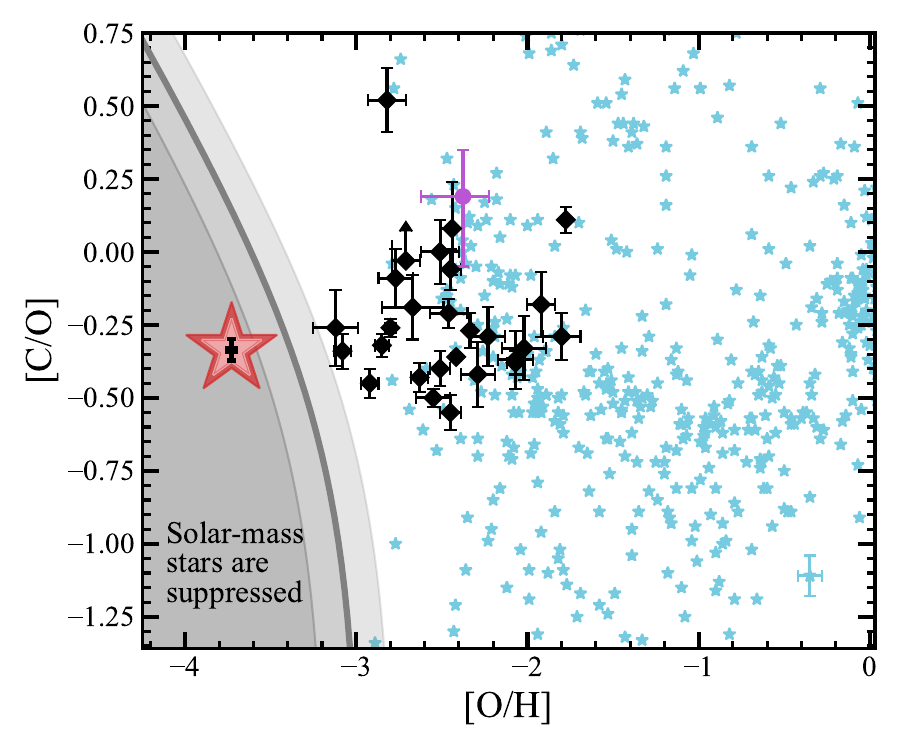}}
\end{center}
\caption{\textbf{The transition discriminant.} With sufficiently low abundances of oxygen and carbon, UMP--DLA--1 (red star) is the first gas reservoir detected below the transition discriminant\cite{Frebel07} (grey area, with uncertainties in light grey) where the IMF is known to be top-heavy. It follows the trend seen in extremely metal-poor and very metal-poor DLAs (black points) of a slight carbon deficit compared to oxygen, while individual metal-poor local stars (blue points, from SAGA\cite{SAGA}) display far more scatter. Characteristic uncertainties for stellar abundances are shown in the bottom-right corner. The purple circle represents the lowest-metallicity galaxy discovered in emission to date\cite{Nakajima25}.
}
\label{fig:forbidden}
\end{figure}

The low C and O metallicities of UMP--DLA--1 make it the first cold gas reservoir to cross into several interesting regimes. 
To first order, the collapse of gaseous mini-halos into star-forming clouds is limited by their ability to dissipate heat arising from adiabatic compressional heating. The assumption that cooling is dominated by fine-structure metal-line cooling of O~{\small{I}} and C~{\small{II}} atoms leads to the so-called ``transition discriminant''\cite{Bromm03,Frebel07}: $D_\text{trans} = \log_{10} \left(10^{\text{[C/H]}} + 0.3 \times 10^\text{[O/H]}\right)$, such that below $D_\text{trans, crit}=-3.5\pm0.2$ the gas cloud will not collapse into stars smaller than $\sim1M_\odot$. Modern hydrodynamical simulations have revealed that cooling due to dust remains efficient at much lower metallicities, and dominates during the actual fragmentation phase of star formation, such that the formation of some stars with $M<1 M_\odot$ cannot be completely prevented \cite{Omukai05,Schneider06}. Nevertheless, the importance of the $D_\text{trans}=-3.5$ threshold is bolstered by other arguments: it is roughly the point at which inefficient cooling significantly slows down collapse by suppressing turbulence\cite{Chon21}, and above $Z\sim 10^{-3.7} Z_\odot$ mini-halo collapse cannot be prevented by any level of Lyman-Werner radiation\cite{BSmith24}. There is a strong consensus that below the discriminant, the IMF is necessarily top-heavy\cite{Chon21,Chiaki22}; this is supported by the fact that old stars in the Milky Way are nearly non-existent in this regime\cite{BSmith24}, as shown in Figure 3.

The fact that UMP--DLA--1 exhibits a $D_\text{trans}=-3.85\pm0.03$ at $z=5.14$ therefore implies that star-formation in ultra-low-metallicity gas, with a top-heavy IMF, remained possible for a long time after Cosmic Dawn, 1.1 billion years after the Big Bang, and even after the completion of reionisation\cite{Bosman22}. 
Its enrichment pattern provides the first direct measurement of the elemental output of \textit{bona fide} SNe from low-metallicity, high-mass progenitors: a slight but highly significant carbon-depletion compared to oxygen, [C/O]$ =-0.34\pm0.03$, and a close alignment of oxygen and silicon, [O/Si]$=0.01\pm0.02$, with respect to their solar abundance ratios.  The relative abundances of these metals, on their own, cannot distinguish between theoretical predictions for the yields of Pop III or low-metallicity Pop II stellar populations \cite{Limongi18,Heger10}. An informative comparison can nevertheless be drawn with extremely-metal-poor DLAs (EMP; $-4<\text{[X/H]}<-3$) which have been discovered at even later times, down to $z\sim3$. EMP DLAs  have very similar Si, O and C abundance ratios as UMP--DLA--1; using measurements of additional metallicities including Al and Fe, previous work has found these patterns to be best explained by enrichment from only a few ($1-30$) Pop III SNe progenitors\cite{Welsh23,Welsh24}.  
Pockets of Pop III and early Pop II star formation, or at least gas reservoirs preserving their enrichment signatures, therefore persisted down to at least $z\sim5$, as suggested by multiple theoretical studies\cite{Trenti09,Liu20,vanVeenen25}. This may imply that, as suggested by these models,  
searching for signs of Pop III stars in remnant pockets of non-enriched gas at $z<7$ may be more fruitful than at $z>20$. %

\begin{figure}[t]
\begin{center}
\resizebox{10cm}{!}{\includegraphics[trim={0 0 0 0},clip]{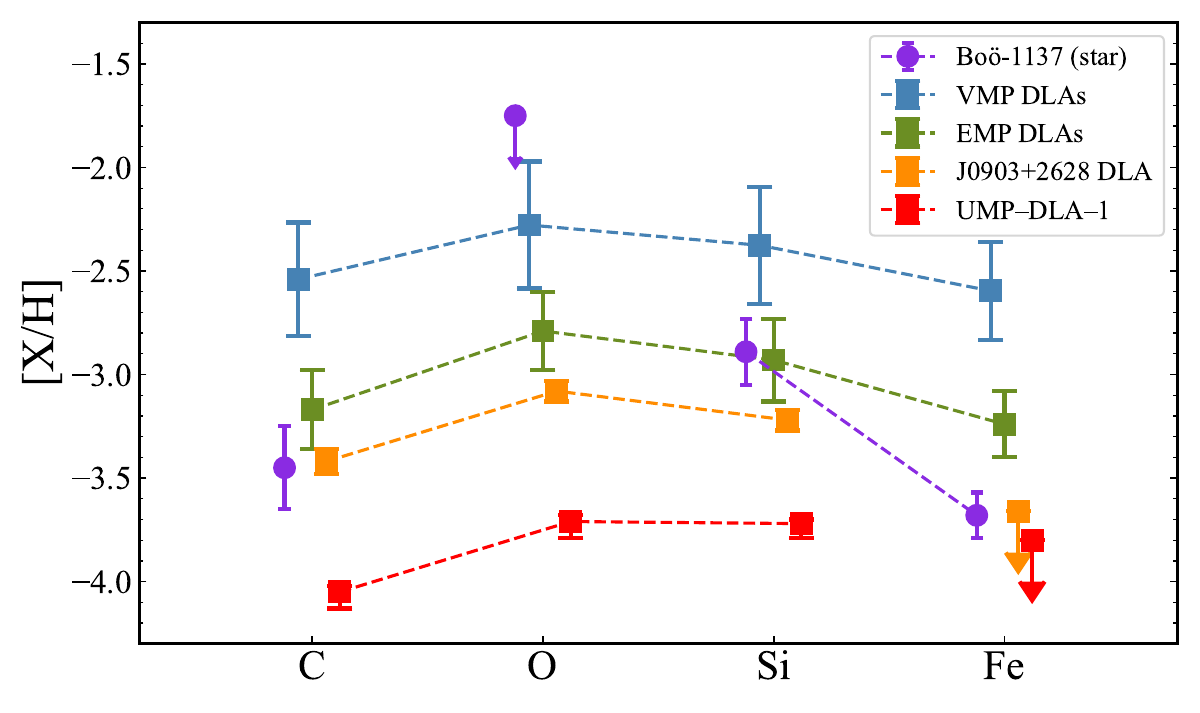}}
\end{center}
\caption{\textbf{Enrichment pattern.} UMP--DLA--1 (red) shows a deficit of carbon compared to $\alpha$ elements (O, Si) which is very similar to that observed in the second-most metal-poor DLA (orange) and in extremely metal-poor (EMP) DLAs. Vertical bars show the full range of measurements within each sample. The same chemical signature is seen in some EMP stars in ultra-faint dwarf galaxies; Bo\"{o}tes-1137 is shown as an example (purple, with vertical bars showing measurement uncertainties). Meanwhile, DLAs which are only very metal-poor do not share the observed carbon-to-oxygen deficit (blue). 
}
\label{fig:compare}
\end{figure}

\smallskip
\textbf{\Large{Implications across cosmic time}}
\smallskip

Comparing UMP--DLA--1's chemical signature to Milky Way stars with measured [O/H] and [C/O] ratios, we find that no star falls below $D_\text{trans}=-3.5\pm0.2$, as shown in Figure 3. However, this is in part a selection effect, as weak oxygen absorption lines corresponding to [O/H]$\lesssim-2.5$ are exceedingly difficult to measure in stellar atmospheres. Two stars are suspected of falling below this value, in both cases due to ultra-metal-poor abundances of iron, carbon and silicon which would require extremely elevated (and probably un-physical) [O/Si] to avoid falling below the discriminant: SDSS J102915.14$+$172927.9 in the Milky Way disc\cite{Caffau11,Lagae23,Caffau24} and SDSS J0715$-$7334, believed to originate from the Large Magellanic Cloud\cite{Limberg24,Ji25}. Although we cannot compare their abundance ratios directly to UMP--DLA--1 as the only metal measured in both is silicon, [Si/Fe] and [Al/Fe] in these stars are consistent with the values found in other EMP DLAs (within the large modeling uncertainties of stellar abundances at low metallicities\cite{Ji25}). In addition, many EMP stars in nearby ultra-faint dwarf galaxies (e.g.~Bo\"{o}tes I/II, Tucana II/III, Coma Berenices, Grus I) have been found to lie on the same trend in [C/H] and [Fe/H] as very metal-poor (VMP) and EMP DLAs\cite{Welsh23}, as shown in Figure 4. Taken together with the late persistence of near-pristine gas reservoirs demonstrated by UMP--DLA--1, it is entirely possible that these local ``relic'' stars are the last remnants of the smallest stars formed within a top-heavy IMF in galaxies similar to UMP--DLA--1. In contrast, the more common ultra iron-poor but carbon-enhanced stars do not fall below the transition discriminant\cite{BSmith24} and do not share the chemical fingerprint of the above stars and DLA populations, and have therefore most likely formed via distinct, albeit exotic pathways\cite{Saccardi23}.

The scarce VMP and EMP DLA populations at $z\sim3$ represent the result of extensive searches through tens of thousands of quasar spectra observed in large surveys. Higher redshift DLAs tend to have lower metallicity\cite{Rafelski12} with tentative indications of more rapid evolution at $z>5$\cite{Rafelski14}. Indeed, of the 5 proximate DLAs where the column of H~{\small{I}} could be measured in the 42 quasars of the E-XQR-30 sample\cite{D'Odorico23,Sodini24}, two are EMP DLAs at $z=6$, hinting at a much higher prevalence of such systems at high redshift. The discovery of UMP--DLA--1 at $z=5.14$ in a blind sample of 30 quasars is nevertheless surprising, and unless our survey was extraordinarily lucky, suggests that more such systems may have eluded identification in existing samples of $z>5$ quasar spectra due to the dense absorption of the $z\gtrsim5$ Ly$\alpha$ forest. If the abundance of such metal-poor DLAs increases towards higher redshift, with their correspondingly weak metal absorption features, they may be an important contaminant for studies of damped H~{\small I} absorption from the intergalactic medium \cite{Simcoe12,Davies25-PDLA}. Future high-resolution spectroscopic surveys with the future generation of extremely large telescopes (e.g.~ELT/ANDES\cite{D'Odorico24}) on the faint quasars to be discovered by \textit{Euclid}\cite{Euclid-quasars} will provide a full census of ultra-metal-poor galaxies and their abundance patterns. 



\section*{Acknowledgements}

SEIB, BS, KP and SH are supported by the Deutsche Forschungsgemeinschaft (DFG) under Emmy Noether grant number BO 5771/1-1. RLD is supported by the Australian Research Council through the Discovery Early Career Researcher Award (DECRA) Fellowship DE240100136 funded by the Australian Government. LK acknowledges the support of a Royal Society University Research Fellowship (grant number URF$\backslash$R1$\backslash$251793).

Based on observations obtained at the international Gemini Observatory, a program of NSF NOIRLab, which is managed by the Association of Universities for Research in Astronomy (AURA) under a cooperative agreement with the U.S. National Science Foundation on behalf of the Gemini Observatory partnership: the U.S. National Science Foundation (United States), National Research Council (Canada), Agencia Nacional de Investigaci\'{o}n y Desarrollo (Chile), Ministerio de Ciencia, Tecnolog\'{i}a e Innovaci\'{o}n (Argentina), Minist\'{e}rio da Ci\^{e}ncia, Tecnologia, Inova\c{c}\~{o}es e Comunica\c{c}\~{o}es (Brazil), and Korea Astronomy and Space Science Institute (Republic of Korea). The data were obtained under Program ID GS-2025A-LP-114, ``The GHOSTLy quasar sample".

\section*{Author contributions statement}

S.~Bosman designed the GHOSTLy observational survey, including the selection of targets which lead to the discovery of UMP--DLA--1. They also designed most of the analysis and drafted most of the manuscript. F.~Davies performed most of the analysis, including optimising the reduction of the data and performing the spectral fitting, and made significant contributions to the manuscript. L.~Welsh provided important insight into the nature of the object and its interpretation. E. Farina helped prepare the observations and provided crucial technical support in their execution. P.~Gaikwad and B.~Spina significantly contributed to drafting the observational proposal, while B.~Maity, K.~Protu\v{s}ov\'{a}, S.~Hwang and L.~Seyberlich made contributions to writing the observing proposal and manuscript. 
The subsequent co-authors (E.~Ba\~{n}ados -- M.~Wolfson) helped write the manuscript, provided comments, and made ongoing contributions as members of the GHOSTLy Collaboration.

\section*{Methods}

\noindent \textbf{Conventions.} Throughout the analysis we assume a flat $\Lambda$CDM cosmology with \textit{Planck}\cite{Planck20} parameters: $H_0=67.36$, $\Omega_\text{M}=0.3153$. To facilitate comparison with nearly all previous literature cited in the main text, we have assumed Solar abundances following Asplund et al.~2009\cite{Asplund09} as a common frame of reference. However, we briefly report here the results obtained by using as a reference the more recent abundance measurements of Magg et al.~2022\cite{Magg22}: UMP--DLA--1 would possess abundances relative to solar of 
[Si/H]$=-3.80_{-0.07}^{+0.02}$, [O/H]$=-3.79_{-0.08}^{+0.03}$, [C/H]$=-4.18_{-0.08}^{+0.03}$ and [Fe/H]$<-3.70 \  (2\sigma)$. In ``real'' terms, our DLA is therefore $\sim16,\!000$ times less carbon-enriched than the Sun. The corresponding relative abundance ratios are [C/O]$=-0.39\pm0.03$ and [O/Si]$=0.01\pm0.02$. The derivation of the critical value of transition discriminant employs the older reference solar abundances of Grevesse \& Sauval 1998\cite{GS98}. In the new standard, the corresponding value would be $D_\text{trans,crit} \sim -3.47$; the difference is therefore much less than the uncertainties involved in the calculation ($\pm 0.1$). The value for UMP--DLA--1 would become $D_\text{trans} = -3.96$, an adjustment of $\sim -0.1$ dex.

\smallskip

\noindent \textbf{Spectroscopic observations.} Precise measurements of DLA metallicities require instruments capable of spectroscopically resolving the width of absorption lines arising from cold gas, corresponding to a minimum resolution $R>30,\!000$. At the same time, increasingly metal-poor DLAs have been found at increasingly early cosmic times (Figure 2), requiring sufficient sensitivity to use fainter quasars as background sources. Within these parameters, the GHOST spectrograph\cite{GHOST1,GHOST2} commissioned in 2024 on the Gemini South Telescope is the most sensitive high-resolution spectrograph built to date. GHOST is a fiber-fed instrument with continuous wavelength coverage of $3480 < \lambda < 10610$\,\AA\ in one fixed setup across two detectors (blue and red arms). We are using GHOST to observe the brightest $30$ quasars at $4.5<z<5.5$ without existing deep high-resolution spectroscopy, as part of the ``GHOSTLy'' large observational programme. 
Among them, the inclusion of the luminous quasar J1941$-$4500 at $z=5.215\pm0.004$\cite{Onken22,Lai24,Schindler20} was serendipitous, although archival data from VLT/X-Shooter\cite{Lai24} already suggested the presence of a metal-poor DLA. 
The background quasar J1941$-$4500 was observed in the standard resolution (SR) mode of the GHOST spectrograph on 25 Jun 2025 ($1\times3000$\,s), 10 Sep 2025 ($3\times3000$\,s), 16 Sep 2025 ($2\times3000$\,s), and 30 Sep 2025 ($1\times3000$\,s), for a total of 21000\,s\,$\simeq$\,5.8\,hr on-source integration time. The spectra were read out in slow mode and binned on-chip by a factor of 8 in the cross-dispersion (cross-fiber) direction to minimize detector read noise. The combined spectrum reached a SNR$\simeq31$ (at $\lambda_\text{rest}=1285$\AA) per $1.85$ km s${}^{-1}$ pixel. 
The effective resolution of the red arm of GHOST in SR mode is $R=57,\!000$\cite{GHOST1}, which we confirmed to be accurate within a few percent from sky emission lines in our data, and this is taken into account in all of the following analysis.

\smallskip

\noindent \textbf{Data reduction.} The Gemini/GHOST data were reduced with a custom version of the public Gemini DRAGONS\cite{DRAGONS1,DRAGONS2} pipeline, plus additional post-processing steps. The unmodified DRAGONS pipeline was used to determine the two-dimensional spectral format, wavelength calibration, and apply a scattered light correction. We implement a custom spectral extraction method, inspired by the PypeIt spectroscopic pipeline\cite{Pypeit-official}, which uses LACosmic\cite{LACosmic} to remove cosmic rays from the two-dimensional echellogram images, then performs optimal extraction\cite{Horne86} with b-spline sky subtraction\cite{Kelson03}. Flux calibration was performed using the PypeIt \texttt{sensfunc} routine applied to an observation of the spectrophotometric standard star EG 21 observed with GHOST in the same mode as the science data. Telluric correction was performed using the PypeIt \texttt{tellfit} routine, which fits a telluric model directly to the science data. The telluric model is built from a principal component analysis decomposition of a large ensemble of theoretical atmospheric transmission spectra computed with the LBLRTM\cite{LBLRTM} code via the TelFit\cite{TelFit} python interface. Finally, the individual order spectra were coadded onto an evenly-spaced wavelength grid with an approximately optimal weighting scheme that includes global signal-to-noise ratio weights, echelle blaze weights, as well as local pixel weights set by the telluric corrections and relative sky background of each exposure.

\smallskip

\noindent \textbf{Continuum modelling and $\boldsymbol{N_\text{HI}}$.} The \textit{a priori} unknown intrinsic quasar continuum is the dominant source of uncertainty in the determination of the H~{\small I} column density of DLAs. Due to its relative proximity to the background quasar, reconstructing the continuum surrounding UMP--DLA--1 is 
challenging due to the strong flux gradient within the quasar's broad Lyman-$\alpha$ emission line. On the other hand, the proximity also increases the number of pixels with high transmission values close to the continuum level. We adopt a parametric Gaussian mixture approach similar to Greig et al.~2017\cite{Greig17}, decomposing the spectral region around rest-frame Lyman-$\alpha$ with a combination of a smooth power-law continuum plus several additive Gaussians describing the detailed shape of the broad emission line features. We vary the number of Gaussians ($n_G$) used to model the Lyman-$\alpha$ emission line  in order to quantitatively account for the impact of this choice. Our fiducial model employs four Gaussians, but we will marginalise over fits which employ three or five Gaussians in order to obtain the final uncertainties on $\nhi$ (see below). In addition, we use 
one Gaussian centered on the rest-frame N~{\small V} $\lambda$1240 line, and a power-law anchored to distant pixels on both the red and blue side of the absorption profile. The spectrum normalised at $\lambda_\text{rest}=1285$ \AA, but not divided by the best-fit quasar continuum, is shown in Extended Data Figure 1.

Fitting the joint model of the DLA and the quasar continuum to the data also requires choosing absorption-free regions 
away from the DLA and LLS
in the spectrum. Even at high spectral resolution $R>50,000$, peaks in the $z=5$ Ly$\alpha$ forest may not commonly reach the continuum level to high precision\cite{Bolton17}. We show the spectral regions used in the fit in Extended Data Figure 2. In addition to the red and blue power-law anchors described above (bottom panel), we include regions without absorption on the red side of the Lyman-$\alpha$ line and over its peak (middle panel). To constrain the DLA absorption, we then employ a short absorbed region in its core and various combinations of the transmission spikes on either side, which we label B1$-$B4 on the blue side and R1$-$R4 on the red side. First, we consider whether these spikes are likely to accurately represent the underlying un-absorbed continuum (except by the DLA and LLS). 

On the blue side, a.k.a.~for the blue power-law (PL) anchors and spikes B1$-$B4, we consider a study by Viel et al.~2013\cite{Viel13} which showed, using hydrodynamical simulations of the intergalactic medium, that the highest peaks in $z=5$ \lal forest reach to $\sim 80\%$ of the continuum, rather than $100\%$, albeit at a somewhat lower resolution ($R\sim40,000$) than our GHOST data ($R\sim57,000$). 
This corrective factor has since often been used both in studies of the \lal power spectrum\cite{Bolton17} as well as when fitting the hydrogen column densities of high-$z$ DLA\cite{Oyarzun25}. To approximately account for this continuum bias, we now multiply the entire spectrum blueward of the DLA trough by a factor of 1.2 for all of our analyses.

On the red side, the Ly$\alpha$ forest absorption in its red wing is suppressed by the extreme ionizing flux from the quasar, i.e.~the proximity effect\cite{Bajtlik88,Davies20-prox}. The hydrogen photoionization rate from the quasar at the location of spike R1, using its absolute magnitude at 1450 \AA, $M_{1450}=-28.2$\cite{Lai24} and assuming a far-UV quasar template\cite{Lusso15}, is $\simeq2.1\times 10^{-12}$\,s$^{-1}$. This is more than triple the value from the ambient UV background, $\Gamma_{\rm HI} \simeq 0.6\times 10^{-12}$\,s$^{-1}$\cite{Gaikwad23, Davies24, Maity26}; in other words, all of the red spikes are well within the quasar's ``proximity zone''. We note that this quasar is among the most luminous known at any redshift, so the magnitude of its proximity effect is larger than the typical quasars observed for DLA analyses. From this enhanced photoionization rate, we expect the mean opacity of the \lal forest to be lower according to $\tau_{\rm eff}\propto\Gamma_{\rm HI}^{-0.55}$\cite{Davies20-prox, Maity26}, resulting in $\tau_{\rm eff} \leq 0.9$ within the region covered by the red spikes. This opacity is analogous to the $z\leq4$ \lal forest, where transmission spikes are far more likely to reach the continuum\cite{Viel13,Bolton17}. We therefore apply no correction to the flux redwards of the DLA, including spikes R1$-$R4.

In order to quantitatively account for the influence of the choice of fitting regions on the resulting $\nhi$, we perform the fit using various combinations of spikes R1$-$R4 and B1$-$B4. Our ``fiducial'' choice is to include spikes B3, B4, R1, R2 and R3, since they appear to be the furthest removed from significant \lal forest absorption (in the subjective opinion of experienced \lal forest astronomers). We also consider a ``minimal'' set of only spikes B3 and R1, and ``maximal'' set using all eight spikes. Finally, we divide the spikes into two independent sets, the ``inner'' set (B3, B4, R1 and R2) and the ``outer'' set (B1, B2, R3 and R4) in order to rule out whether any individual feature is driving the $\nhi$ measurement.

We perform fifteen different fits, corresponding to all permutations of the three choices for $n_G$ and five choices of the included fitting regions. For each combination of choices, we fit for the H~{\small I} column density of UMP--DLA--1 jointly with the 15$-$21 parameters of the continuum described above -- three for each Gaussian, plus two for the power-law component -- as well as the $N_{\rm HI}$ of the proximate LLS and its Doppler parameter $b_\text{LLS}$. 
Note that the redshifts of the DLA and LLS are known accurately from fitting of their metal absorption (see next Methods sub-section) and are therefore not varied during the $\nhi$ fits. 
We obtain posterior PDFs from a Markov chain Monte Carlo (MCMC) analysis. 
A prior requires the Gaussian components to be centered at increasing wavelengths, i.e.~Gaussian number 1 must be centered at a lower wavelength than Gaussian number 2, etc. This serves to stop the MCMC from exploring ``swaps'' of the Gaussians where the same shape is be obtained by swapping the parameter values of two Gaussians. In addition, each Gaussian is given very wide uniform priors of $7400<\lambda_C<7700$\AA \ and $200<\sigma<10000$ km s$^{-1}$, where $\lambda_C$ is the central wavelength and $\sigma$ is the standard deviation in velocity space. The amplitude is bound to be positive, with a flat prior from zero to the peak value of the spectrum. 
The resulting full corner plots are impractically large to be included in the text, so we instead make them available, for all fifteen fits, as online-only additional material (Supplementary Information).  We find that, for all models, all parameters are well-constrained, with the sole exception of $b_\text{LLS}$ which is always unconstrained within its uniform prior range $0< b_\text{LLS} \leq 25$ km s$^{-1}$.

The marginalised posteriors for $\nhi$ from all fifteen fits are shown in Extended Data Figure 3. Most posteriors agree within $1\sigma$, except for those where only three Gaussians are employed to model the \lal emission line, which prefer somewhat higher values of $\nhi$. We note that this comes at the cost of significantly worsened fits ($\Delta \chi^2 > 120$ compared to the same models with four Gaussians), suggesting that three components are not sufficient to model the line. Nevertheless, to be conservative, we employ the full $1\sigma$ envelope of all fifteen fits as the uncertainty of $\nhi$ in our analysis. We use the marginalised central value of $\nhi$ from our fiducial fit as the reference value, since our analysis demonstrates that the precise choice of $n_G$ and fitting regions has only a small impact on the final $\nhi$ measurement. This results in our final measurement of $\log \nhi/\text{cm}^{-2} = 20.542_{-0.014}^{+0.071}$. As a final check, we re-ran the fitting procedure without including any of the spikes on either side of the DLA. In that case, the fit becomes unconstrained for $\nhi$. This supports the idea that our parameterisation of the continuum and the fitting priors are not intrinsically  favouring a large $\nhi$.

As an independent, qualitative test of the high $N_{\rm HI}$ of UMP--DLA--1, we inspected higher order Lyman series lines in the GHOST spectrum. Due to its relatively high redshift, the Lyman series forests are strongly absorbed, but any detection of transmission would still rule out the DLA interpretation of the Ly$\alpha$ absorption feature. No such signal is detected for any Lyman series transition, as shown in Extended Data Figure 4 for the first eight higher order transitions (Ly$\beta$ through Ly$-9$, labeled as Ly$n$ where $n$ is their order in the Lyman series). The extent of the saturated trough is consistent with a Doppler parameter of $b\simeq12.5$\,km/s, corresponding to a thermal broadening of $\sim8000$\,K. 
However, due to the substantial contamination by unrelated foreground absorption at these wavelengths this can only be treated as an upper limit. We note that the length of the absorbed troughs in the lower-order Lyman series are increased, as expected, by the presence of deuterium. Extended Figure 4 shows the expected absorption assuming a primordial D/H abundance\cite{Cooke11}.

\smallskip

\noindent \textbf{Metal absorption systems.} Column densities of Si~{\small II}, O~{\small I}, C~{\small II}, and Fe~{\small II} were fit to $\pm100$\,km/s segments of the spectrum centered around their corresponding absorption features. We assumed that all of the absorption lines are described by a single kinematic component modeled by a Voigt profile with a Doppler parameter composed of both turbulent and thermal components, $b=\sqrt{b_{\rm turb}^2+b_{\rm therm}^2}$, at a single redshift $z_{\rm abs}$. 
Some metal absorbers in DLAs are found to have Doppler parameters smaller than our spectral resolution, and it is therefore possible that our system is composed of multiple sub-components\cite{Jorgenson09}. Nevertheless, the lack of detection of Si~{\small{II}} 1304\AA, as shown in Fig.~\ref{fig:spec}, 
demonstrates that the low-ionisation Si absorption cannot be saturated; the column densities measured assuming a single component are therefore accurately tracing the total metal content of the system. We also point out that a substantial fraction of the most metal-poor DLAs known show a simple kinematic structure: 2/7 of VMP DLAs\cite{Cooke11} were well-modeled by a single component, and this fraction seems to increase with decreasing metallicity, with 5/8 of the EMP DLAs displaying a single absorption component\cite{Welsh24}. 
For each line, we model the local continuum with an independent power-law, and fit the continuum jointly with the absorption. The resulting MCMC constraints on the column densities, $b_{\rm turb}$, and gas temperature $T$, after marginalizing over the continuum parameters, are shown in Extended Data Figure 5. While $N_{\rm SiII}$, $N_{\rm OI}$, $N_{\rm CII}$, and $b_{\rm turb}$ are well-constrained, $N_{\rm FeII}$ and $T$ are not. The absorption lines of Fe~{\small II} are undetected, and we do not detect any dependence of the absorption line widths with the atomic weights of the corresponding elements. Upper limits on $N_{\rm FeII}$ and $T$ are conservatively determined from the posterior PDFs by recording the value at which the PDF falls by a factor of $e^{-2}$ from its peak (i.e.~$\Delta \chi^2 = 4$). 

We show the posterior PDFs for the relative abundances of silicon, oxygen, carbon and iron in Extended Data Figure 5, combining posterior samples of $N_{\rm HI}$ with the corresponding low-ionization metal columns and adopting the Asplund et al.~2009\cite{Asplund09} abundances.

We obtain relative abundances of silicon, oxygen, carbon and iron by combining the uncertainties from the fitting of the metal absorption systems and the fitting of $\nhi$. 
Note that, due to the relatively high $N_{\rm HI}$ of UMP--DLA--1, these abundances are not subject to any significant ionization corrections. That is, the elemental abundances are very nearly equal to the abundances of the dominant low-ionization species that we observe, unlike lower $N_{\rm HI}$ absorbers such as Lyman limit systems where the significant corrections required to account for the unknown ionization state of the gas introduce large systematic uncertainties\cite{Fumagalli11}. 
This is also in contrast to the situation for stars, where often highly sophisticated modeling of three-dimensional non-local-thermodynamical-equilibrium (3D NLTE) effects is required to avoid substantial systematic measurement bias\cite{Bergemann25}.

Due to the relatively high redshift of UMP--DLA--1, additional absorption features commonly used in metal-poor DLA abundance measurements (e.g.~table~3 in Cooke et al. 2011\cite{Cooke11}) that lie blueward of Lyman-$\alpha$ ($\lambda_{\rm rest}<1215.67$\,\AA) are too contaminated by the foreground Lyman-$\alpha$ forest for robust quantitative analysis. No clear signal is seen for any additional absorption line except for C~{\small II} $\lambda$1036, which fortuitously lands in a relatively transmissive region inside the Lyman-$\beta$ proximity zone of the background quasar; due to the remaining systematic uncertainty in the shape of the contaminated continuum, we do not include this line in the fitting above, but we note that our best-fit $N_{\rm CII}$ and $b_{\rm turb}$ is fully consistent with the observed absorption line. 

\smallskip

\noindent \textbf{The Lyman-limit system.} Given the location of UMP--DLA--1 within relative proximity of a bright a quasar (although it is outside the proximity zone), it is reasonable to ask whether the presence of a LLS in-between the DLA and the quasar is providing shielding which helps maintain a high neutral fraction in the DLA. However, as we detail below, the DLA would still be a DLA in the absence of the LLS.
Computing the hydrogen photoionization rate from the quasar at the location of the DLA in the same way as above, we find $\Gamma_{\rm HI}=1.1\times$\,10${}^{-12}$\,s$^{-1}$ in the absence of the LLS. This is about twice the value of the UV background at $z=5.15$\cite{Gaikwad23,Davies24}, meaning that the presence of the LLS has decreased the strength of the radiation onto the DLA by a factor of $\sim2$. 
We approximate the effect of this enhanced ionisation on the H~{\small{I}} content of the DLA by computing a simple slab model at constant gas density and temperature photoionized by an external background. We assume $T=8000$\,K and set the width of the slab to match the observed $N_{\rm HI}$ of UMP--DLA--1 assuming no quasar contribution to the UV background, then re-compute the ionisation profiles with the quasar ``switched on.'' Only for very low hydrogen number densities of $n_{\rm H}<0.01$\,cm$^{-3}$, illumination by the quasar would reduce the $N_{\rm HI}$ below the DLA threshold of $2\times10^{20}$\,cm$^{-2}$. However, very metal-poor DLAs ([X/H]$\lesssim-2$) of comparable $N_{\rm HI}$ have typical inferred $n_{\rm H}$ of $\sim0.1$\,cm$^{-3}$\cite{Cooke15}, where the impact of the LLS shadowing on $N_{\rm HI}$ would be very small ($<0.1$\,dex). We cannot measure $n_\text{H}$ directly in our system due to the lack of higher-ionisation metal absorption $-$ since $n_\text{H}$ is measured by using ratio of elements in different ionisation states. However, it is difficult to conceive of a physical reason why the physical density of DLAs would drop abruptly by factors more than $10$ as their metallicities are lowered. 
Therefore, we conclude that unless UMP$-$DLA$-$1 has an unexpectedly low $n_\text{H}$ by a factor of $\gtrsim10$, the presence of the LLS is very unlikely to affect its self-shielded or ``DLA status''.
\smallskip

\noindent \textbf{Literature compilation.} The literature compilation of DLAs in Figure 2 consists of 183 objects with a previous compilation as the starting basis\cite{Rafelski12}, from which we use metallicities reported in 45 previous studies\cite{Wolfe94,Lu96,Prochaska96,Prochaska97a,Lu98,Lopez99,Pettini99, Prochaska99, Molaro00,Petitjean00,Pettini00,Prochaska00,Srianand00,Dessauges01,Ellison01, Molaro01, Prochaska01a, Prochaska01b,Ledoux02,Levshakov02,Lopez02,Petitjean02,Prochaska02, Centurion03, Ledoux03, Lopez03, Prochaska03a,Prochaska03b,Dessauges04,Khare04,Turnshek04, Akerman05,Rao05,Dessauges06,Ledoux06,Meiring06,Dessauges07,Ellison07,Meiring07, Prochaska07, Noterdaeme08,Wolfe08, Jorgenson10,Meiring11,Vladilo11}. Newer DLA measurements at higher redshifts are added from more recent studies\cite{Cooke11,Rafelski14,Dutta14,Cooke15,Wang15,Cooke17,Banados19,Sodini24}. In general, carbon abundances are reported less commonly than those of more reliable tracers of $\alpha$-element abundances (Si, O), explaining the lower number of DLAs included in Figure 3. We did not re-measure abundances from any archival spectra for the literature compilation; some of the oldest measurements may be assuming different references for solar abundances leading to extra systematic uncertainties up to $\sim0.3$ on [$\alpha$/H]. In contrast, all measurements of [O/H] and [C/O] in Figure 3 use the same reference\cite{Asplund09}.

\smallskip

\noindent \textbf{Data availability.} The raw observational data underlying this paper, and all associated calibration observations, will become publicly available on the Gemini Observatory Archive after the standard proprietary period of twelve months (June$-$September 2026). The science-grade spectrum reduced using our custom pipeline will be made available upon reasonable request to the authors.

\smallskip

\noindent \textbf{Code availability.} Gemini Observatory provides a data-reduction pipeline for the GHOST instrument, DRAGONS, which is public\cite{DRAGONS2}. Our custom reduction pipeline will be rendered public at the conclusion of the currently ongoing observations of quasars as part of the GHOSTLy program, in 2027 at the latest.

\section*{Extended Data Figures}

\captionsetup{labelformat=empty,labelsep=none}

\begin{figure}[htb]
\begin{center}
\resizebox{14cm}{!}{\includegraphics[trim={0 0 0 0},clip]{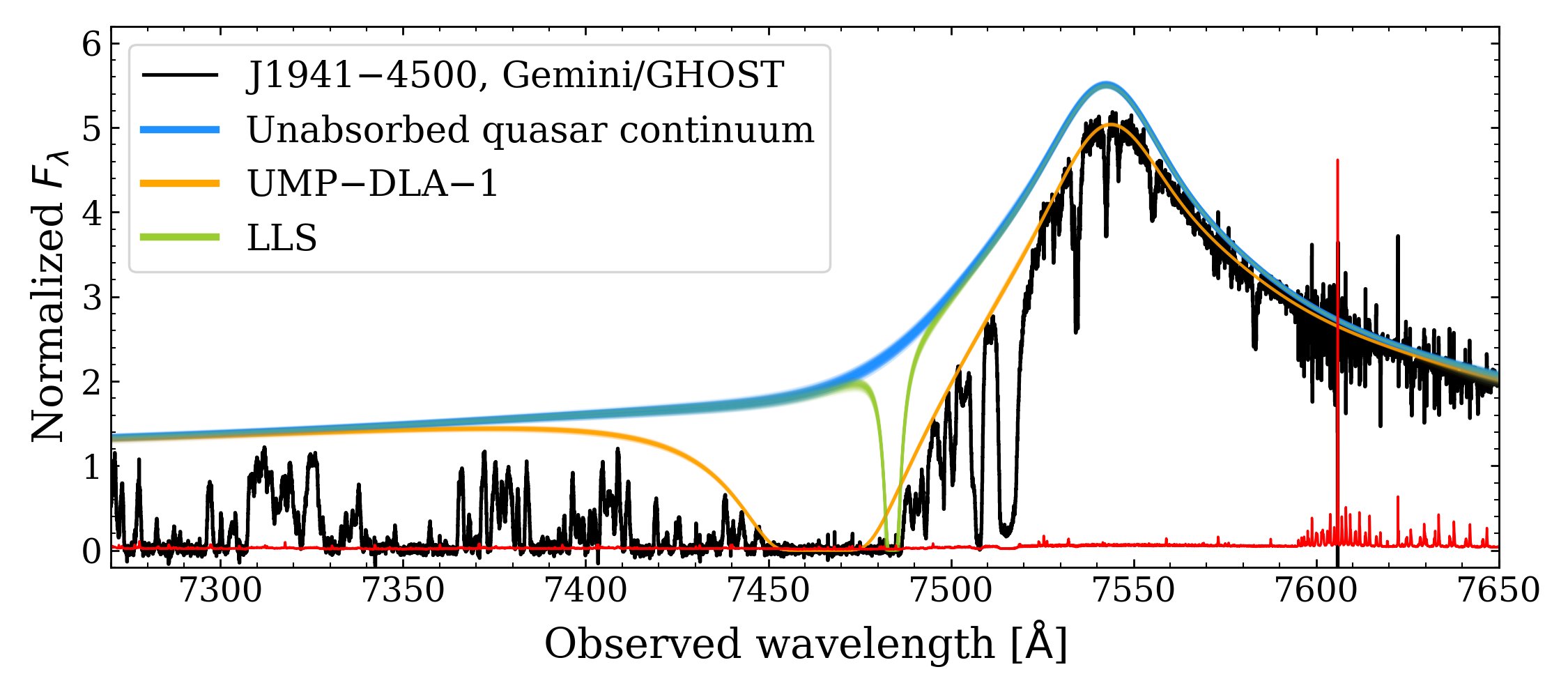}}
\end{center}
\caption{\textbf{Extended Data Figure 1.} Same as panel (a) of Figure 1, but showing the un-normalised continuum with 100 draws from the 17-parameter quasar continuum model (blue) and the corresponding 100 draws of the DLA (orange) and LLS (green) parameters. The flux is normalised such that $F_\lambda = 1$ represents the flux at $\lambda_\text{rest} = 1285$\AA.}
\label{fig:unnorm}
\end{figure}

\begin{figure}
\begin{center}
\resizebox{17cm}{!}{\includegraphics[trim={0 0 0 0},clip]{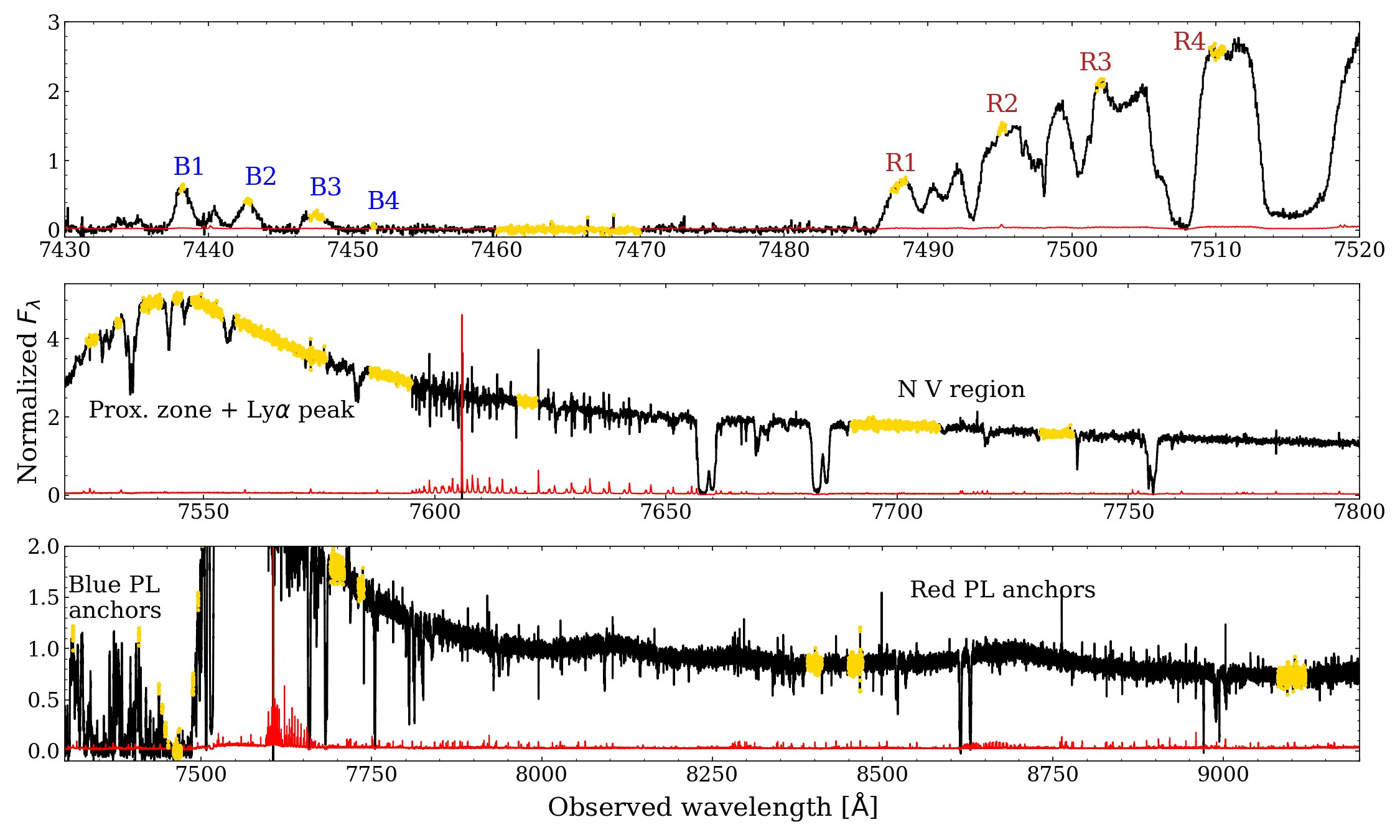}}
\end{center}
\caption{\textbf{Extended Data Figure 2.} Regions used in the fitting of the $\nhi$ of the DLA and LLS  are highlighted in yellow and labeled as described in the Methods. Black shows spectral ranges not included in the fits, with corresponding uncertainties in red. The spectrum has been normalised such at $F_\lambda=1$ represents the flux at $\lambda_\text{rest}=1285$\AA.}
\label{fig:regions}
\end{figure}

\begin{figure}[htb]
\begin{center}
\resizebox{10cm}{!}{\includegraphics[trim={0 0 0 0},clip]{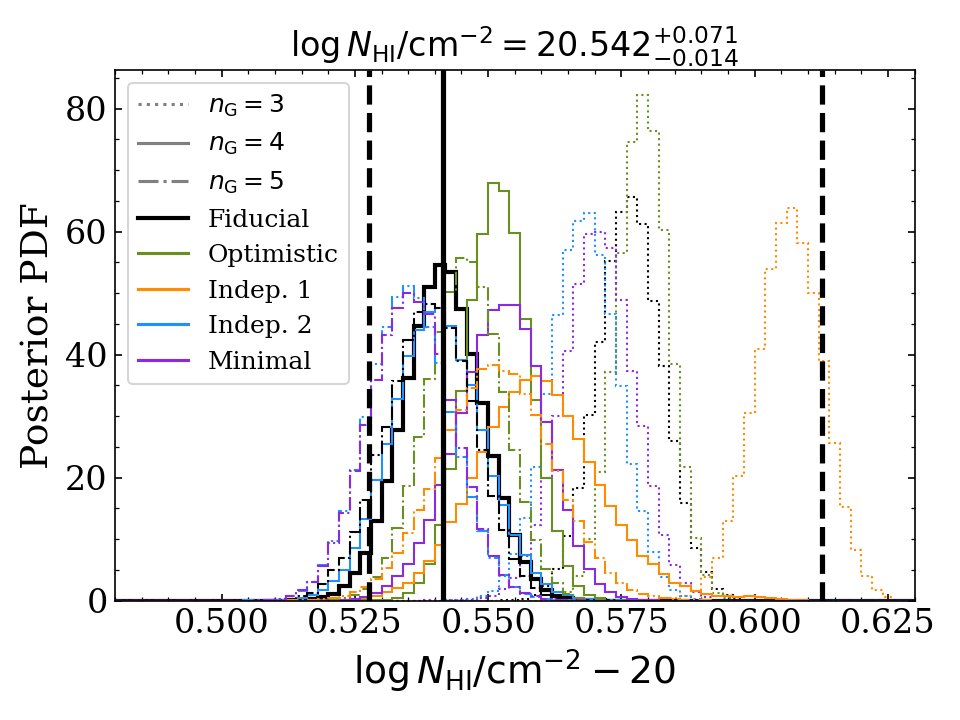}}
\end{center}
\caption{\textbf{Extended Data Figure 3.} Comparison of the marginalised posterior distributions of $\nhi$ for all fifteen permutations of fitting choices $-$ three choices for the number of Gaussians representing the \lal emission line, and five choices corresponding to different combinations of the fitting regions B1$-$B4 and R1$-$R4. The full envelope of the $1\sigma$ uncertaintites of the individual fits gives us the total uncertaintites on the fiducial $\nhi$.}
\label{fig:forbidden}
\end{figure}

\begin{figure}[htb]
\begin{center}
\resizebox{14cm}{!}{\includegraphics[trim={0 0 0 0},clip]{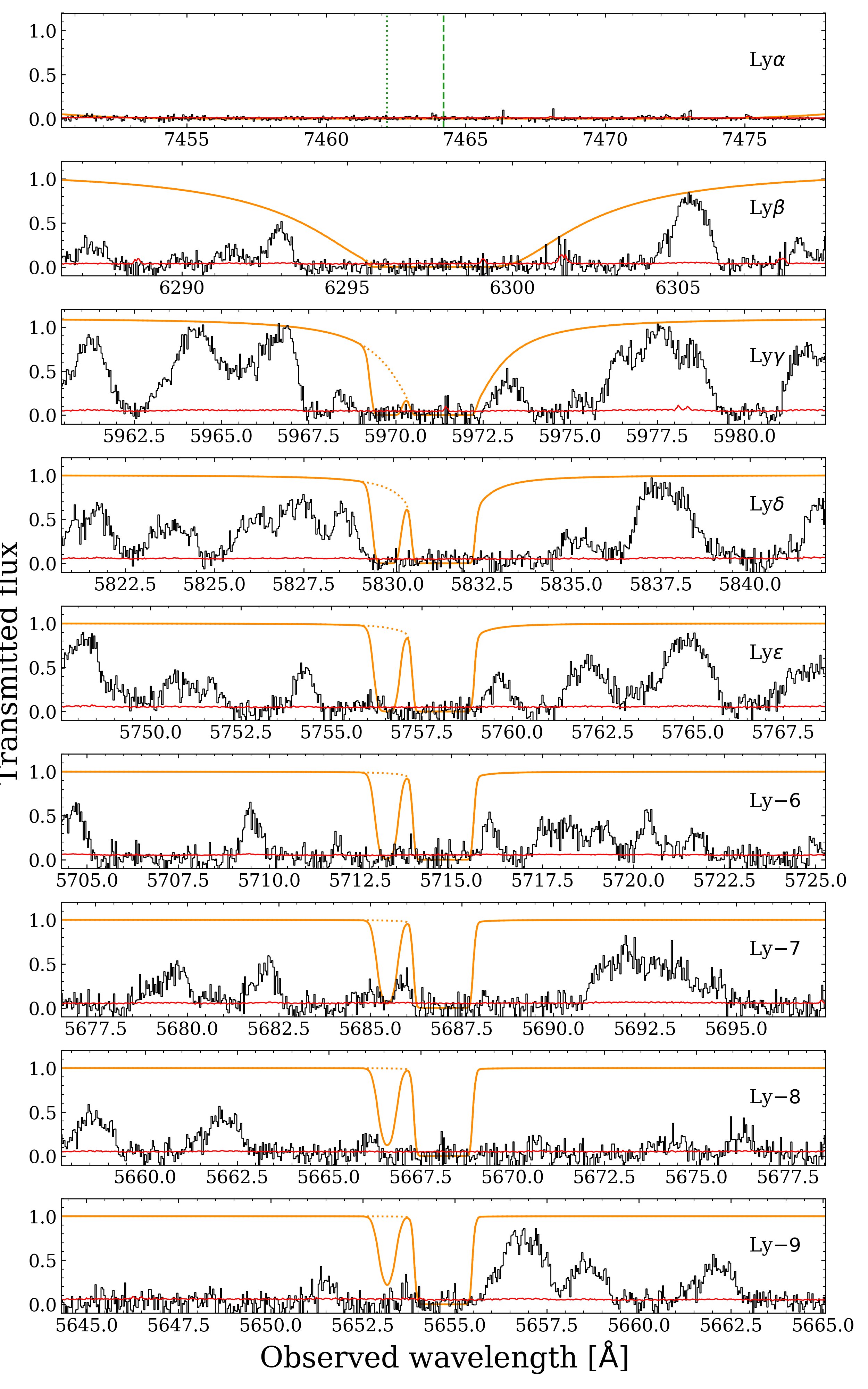}}
\end{center}
\vskip-2em
\caption{\textbf{Extended Data Figure 4.} Spectrum of the background quasar (black) and best-fit H~{\small{I}} absorption profiles obtained from the Lyman-$\alpha$ absorption profile (orange). All transitions of the Lyman-series are fully opaque at the redshift of UMP--DLA--1, confirming its high hydrogen column density. The displayed velocity range is $\pm 550$ km s${}^{-1}$. Absorption by deuterium is also shown as the left-most component of the orange lines in all panels.
}
\label{fig:forbidden}
\end{figure}

\begin{figure}[htb]
\begin{center}
\resizebox{17cm}{!}{\includegraphics[trim={0 0 0 0},clip]{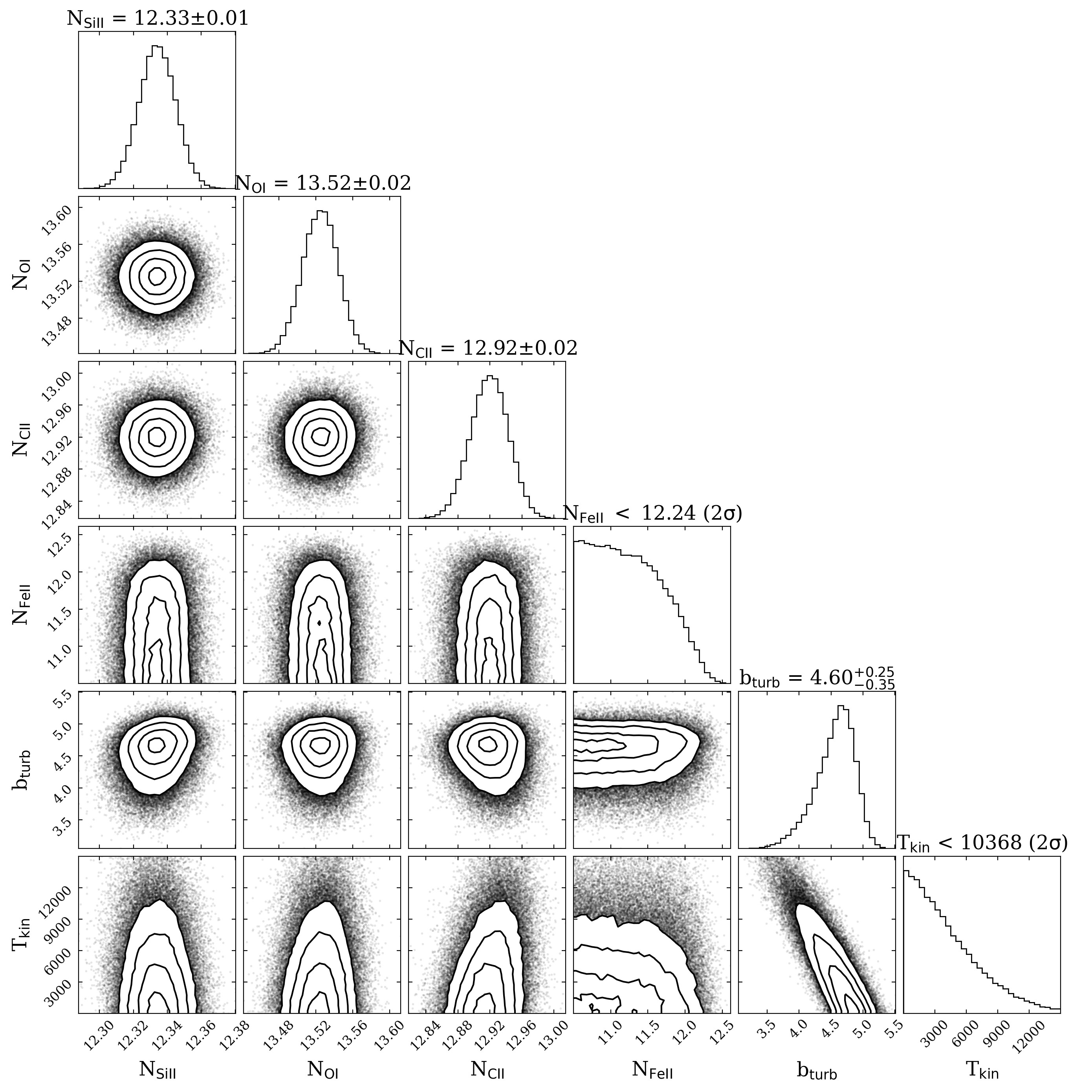}}
\end{center}
\caption{\textbf{Extended Data Figure 5.} Marginalised posterior of the column densities of all detected metal absorption systems and the line-broadening parameters $b_\text{turb}$ and $T$. Continuum parameters have been marginalised over.}
\label{fig:forbidden}
\end{figure}


\begin{thebibliography}{100}
\urlstyle{rm}
\expandafter\ifx\csname url\endcsname\relax
  \def\url#1{\texttt{#1}}\fi
\expandafter\ifx\csname urlprefix\endcsname\relax\def\urlprefix{URL }\fi
\expandafter\ifx\csname doiprefix\endcsname\relax\def\doiprefix{DOI: }\fi
\providecommand{\bibinfo}[2]{#2}
\providecommand{\eprint}[2][]{\url{#2}}

\bibitem{Bromm03}
\bibinfo{author}{{Bromm}, V.} \& \bibinfo{author}{{Loeb}, A.}
\newblock \bibinfo{journal}{\bibinfo{title}{{The formation of the first
  low-mass stars from gas with low carbon and oxygen abundances}}}.
\newblock {\emph{\JournalTitle{\nat}}} \textbf{\bibinfo{volume}{425}},
  \bibinfo{pages}{812--814}, \doiprefix\url{10.1038/nature02071}
  (\bibinfo{year}{2003}).
\newblock \eprint{astro-ph/0310622}.

\bibitem{Frebel07}
\bibinfo{author}{{Frebel}, A.}, \bibinfo{author}{{Johnson}, J.~L.} \&
  \bibinfo{author}{{Bromm}, V.}
\newblock \bibinfo{journal}{\bibinfo{title}{{Probing the formation of the first
  low-mass stars with stellar archaeology}}}.
\newblock {\emph{\JournalTitle{\mnras}}} \textbf{\bibinfo{volume}{380}},
  \bibinfo{pages}{L40--L44}, \doiprefix\url{10.1111/j.1745-3933.2007.00344.x}
  (\bibinfo{year}{2007}).
\newblock \eprint{astro-ph/0701395}.

\bibitem{GHOST1}
\bibinfo{author}{{McConnachie}, A.~W.} \emph{et~al.}
\newblock \bibinfo{journal}{\bibinfo{title}{{The Science Performance of the
  Gemini High Resolution Optical Spectrograph}}}.
\newblock {\emph{\JournalTitle{\pasp}}} \textbf{\bibinfo{volume}{136}},
  \bibinfo{pages}{035001}, \doiprefix\url{10.1088/1538-3873/ad1ed4}
  (\bibinfo{year}{2024}).
\newblock \eprint{2401.07452}.

\bibitem{GHOST2}
\bibinfo{author}{{Kalari}, V.~M.} \emph{et~al.}
\newblock \bibinfo{journal}{\bibinfo{title}{{Gemini High-resolution Optical
  Spectrograph (GHOST) at Gemini South: Instrument Performance and Integration,
  First Science, and Next Steps}}}.
\newblock {\emph{\JournalTitle{\aj}}} \textbf{\bibinfo{volume}{168}},
  \bibinfo{pages}{208}, \doiprefix\url{10.3847/1538-3881/ad72ed}
  (\bibinfo{year}{2024}).
\newblock \eprint{2409.05855}.

\bibitem{Welsh23}
\bibinfo{author}{{Welsh}, L.}, \bibinfo{author}{{Cooke}, R.},
  \bibinfo{author}{{Fumagalli}, M.} \& \bibinfo{author}{{Pettini}, M.}
\newblock \bibinfo{journal}{\bibinfo{title}{{Towards ultra metal-poor DLAs:
  linking the chemistry of the most metal-poor DLA to the first stars}}}.
\newblock {\emph{\JournalTitle{\mnras}}} \textbf{\bibinfo{volume}{525}},
  \bibinfo{pages}{527--541}, \doiprefix\url{10.1093/mnras/stad2181}
  (\bibinfo{year}{2023}).
\newblock \eprint{2307.03771}.

\bibitem{Bromm11}
\bibinfo{author}{{Bromm}, V.} \& \bibinfo{author}{{Yoshida}, N.}
\newblock \bibinfo{journal}{\bibinfo{title}{{The First Galaxies}}}.
\newblock {\emph{\JournalTitle{\araa}}} \textbf{\bibinfo{volume}{49}},
  \bibinfo{pages}{373--407},
  \doiprefix\url{10.1146/annurev-astro-081710-102608} (\bibinfo{year}{2011}).
\newblock \eprint{1102.4638}.

\bibitem{Marks12}
\bibinfo{author}{{Marks}, M.}, \bibinfo{author}{{Kroupa}, P.},
  \bibinfo{author}{{Dabringhausen}, J.} \& \bibinfo{author}{{Pawlowski}, M.~S.}
\newblock \bibinfo{journal}{\bibinfo{title}{{Evidence for top-heavy stellar
  initial mass functions with increasing density and decreasing metallicity}}}.
\newblock {\emph{\JournalTitle{\mnras}}} \textbf{\bibinfo{volume}{422}},
  \bibinfo{pages}{2246--2254}, \doiprefix\url{10.1111/j.1365-2966.2012.20767.x}
  (\bibinfo{year}{2012}).
\newblock \eprint{1202.4755}.

\bibitem{Santoro06}
\bibinfo{author}{{Santoro}, F.} \& \bibinfo{author}{{Shull}, J.~M.}
\newblock \bibinfo{journal}{\bibinfo{title}{{Critical Metallicity and
  Fine-Structure Emission of Primordial Gas Enriched by the First Stars}}}.
\newblock {\emph{\JournalTitle{\apj}}} \textbf{\bibinfo{volume}{643}},
  \bibinfo{pages}{26--37}, \doiprefix\url{10.1086/501518}
  (\bibinfo{year}{2006}).
\newblock \eprint{astro-ph/0509101}.

\bibitem{Klessen23}
\bibinfo{author}{{Klessen}, R.~S.} \& \bibinfo{author}{{Glover}, S. C.~O.}
\newblock \bibinfo{journal}{\bibinfo{title}{{The First Stars: Formation,
  Properties, and Impact}}}.
\newblock {\emph{\JournalTitle{\araa}}} \textbf{\bibinfo{volume}{61}},
  \bibinfo{pages}{65--130}, \doiprefix\url{10.1146/annurev-astro-071221-053453}
  (\bibinfo{year}{2023}).
\newblock \eprint{2303.12500}.

\bibitem{Nakajima25}
\bibinfo{author}{{Nakajima}, K.} \emph{et~al.}
\newblock \bibinfo{journal}{\bibinfo{title}{{An ultra-faint, chemically
  primitive galaxy forming in the reionization era}}}.
\newblock {\emph{\JournalTitle{\nat}}} \textbf{\bibinfo{volume}{653}},
  \bibinfo{pages}{363--367}, \doiprefix\url{10.1038/s41586-026-10374-1}
  (\bibinfo{year}{2026}).
\newblock \eprint{2506.11846}.

\bibitem{Cullen25}
\bibinfo{author}{{Cullen}, F.} \emph{et~al.}
\newblock \bibinfo{journal}{\bibinfo{title}{{The JWST EXCELS survey: an
  extremely metal-poor galaxy at z = 8.271 hosting an unusual population of
  massive stars}}}.
\newblock {\emph{\JournalTitle{\mnras}}} \textbf{\bibinfo{volume}{540}},
  \bibinfo{pages}{2176--2194}, \doiprefix\url{10.1093/mnras/staf838}
  (\bibinfo{year}{2025}).
\newblock \eprint{2501.11099}.

\bibitem{Wolfe86}
\bibinfo{author}{{Wolfe}, A.~M.}, \bibinfo{author}{{Turnshek}, D.~A.},
  \bibinfo{author}{{Smith}, H.~E.} \& \bibinfo{author}{{Cohen}, R.~D.}
\newblock \bibinfo{journal}{\bibinfo{title}{{Damped Lyman-alpha absorption by
  disk galaxies with large redshifts. I - The Lick survey}}}.
\newblock {\emph{\JournalTitle{\apjs}}} \textbf{\bibinfo{volume}{61}},
  \bibinfo{pages}{249--304}, \doiprefix\url{10.1086/191114}
  (\bibinfo{year}{1986}).

\bibitem{Krogager17}
\bibinfo{author}{{Krogager}, J.-K.}, \bibinfo{author}{{M{\o}ller}, P.},
  \bibinfo{author}{{Fynbo}, J.~P.~U.} \& \bibinfo{author}{{Noterdaeme}, P.}
\newblock \bibinfo{journal}{\bibinfo{title}{{Consensus report on 25 yr of
  searches for damped Ly {\ensuremath{\alpha}} galaxies in emission: confirming
  their metallicity-luminosity relation at z {\ensuremath{\gtrsim}} 2}}}.
\newblock {\emph{\JournalTitle{\mnras}}} \textbf{\bibinfo{volume}{469}},
  \bibinfo{pages}{2959--2981}, \doiprefix\url{10.1093/mnras/stx1011}
  (\bibinfo{year}{2017}).
\newblock \eprint{1704.08075}.

\bibitem{Neeleman17}
\bibinfo{author}{{Neeleman}, M.} \emph{et~al.}
\newblock \bibinfo{journal}{\bibinfo{title}{{[C II] 158-{\ensuremath{\mu}}m
  emission from the host galaxies of damped Lyman-alpha systems}}}.
\newblock {\emph{\JournalTitle{Science}}} \textbf{\bibinfo{volume}{355}},
  \bibinfo{pages}{1285--1288}, \doiprefix\url{10.1126/science.aal1737}
  (\bibinfo{year}{2017}).
\newblock \eprint{1703.07797}.

\bibitem{Neeleman25}
\bibinfo{author}{{Neeleman}, M.}, \bibinfo{author}{{Kanekar}, N.},
  \bibinfo{author}{{Prochaska}, J.~X.}, \bibinfo{author}{{Rafelski}, M.~A.} \&
  \bibinfo{author}{{Kahinga}, L.~A.}
\newblock \bibinfo{journal}{\bibinfo{title}{{A [C II] 158 {\ensuremath{\mu}}m
  Survey of Damped Ly{\ensuremath{\alpha}} Absorber Galaxies at z
  {\ensuremath{\sim}} 4: Observations of Dense and Metal-enriched Neutral Gas
  within the Circumgalactic Medium of Star-forming Galaxies}}}.
\newblock {\emph{\JournalTitle{\apj}}} \textbf{\bibinfo{volume}{983}},
  \bibinfo{pages}{26}, \doiprefix\url{10.3847/1538-4357/adbbd1}
  (\bibinfo{year}{2025}).
\newblock \eprint{2504.03828}.

\bibitem{Beaver72}
\bibinfo{author}{{Beaver}, E.~A.}, \bibinfo{author}{{Burbidge}, E.~M.},
  \bibinfo{author}{{McIlwain}, C.~E.}, \bibinfo{author}{{Epps}, H.~W.} \&
  \bibinfo{author}{{Strittmatter}, P.~A.}
\newblock \bibinfo{journal}{\bibinfo{title}{{Digicon Spectrophotometry of the
  Quasi-Stellar Object PHL 957}}}.
\newblock {\emph{\JournalTitle{\apj}}} \textbf{\bibinfo{volume}{178}},
  \bibinfo{pages}{95--104}, \doiprefix\url{10.1086/151768}
  (\bibinfo{year}{1972}).

\bibitem{Cooke11}
\bibinfo{author}{{Cooke}, R.}, \bibinfo{author}{{Pettini}, M.},
  \bibinfo{author}{{Steidel}, C.~C.}, \bibinfo{author}{{Rudie}, G.~C.} \&
  \bibinfo{author}{{Nissen}, P.~E.}
\newblock \bibinfo{journal}{\bibinfo{title}{{The most metal-poor damped
  Ly{\ensuremath{\alpha}} systems: insights into chemical evolution in the very
  metal-poor regime}}}.
\newblock {\emph{\JournalTitle{\mnras}}} \textbf{\bibinfo{volume}{417}},
  \bibinfo{pages}{1534--1558}, \doiprefix\url{10.1111/j.1365-2966.2011.19365.x}
  (\bibinfo{year}{2011}).
\newblock \eprint{1106.2805}.

\bibitem{Schindler20}
\bibinfo{author}{{Schindler}, J.-T.} \emph{et~al.}
\newblock \bibinfo{journal}{\bibinfo{title}{{The X-SHOOTER/ALMA Sample of
  Quasars in the Epoch of Reionization. I. NIR Spectral Modeling, Iron
  Enrichment, and Broad Emission Line Properties}}}.
\newblock {\emph{\JournalTitle{\apj}}} \textbf{\bibinfo{volume}{905}},
  \bibinfo{pages}{51}, \doiprefix\url{10.3847/1538-4357/abc2d7}
  (\bibinfo{year}{2020}).
\newblock \eprint{2010.06902}.

\bibitem{Ellison02}
\bibinfo{author}{{Ellison}, S.~L.} \emph{et~al.}
\newblock \bibinfo{journal}{\bibinfo{title}{{The CORALS survey. II. Clues to
  galaxy clustering around QSOs from z$_{abs}$ \raisebox{-0.5ex}\textasciitilde
  z$_{em}$ damped Lyman alpha systems}}}.
\newblock {\emph{\JournalTitle{\aap}}} \textbf{\bibinfo{volume}{383}},
  \bibinfo{pages}{91--97}, \doiprefix\url{10.1051/0004-6361:20011738}
  (\bibinfo{year}{2002}).
\newblock \eprint{astro-ph/0112135}.

\bibitem{BeersChristlieb05}
\bibinfo{author}{{Beers}, T.~C.} \& \bibinfo{author}{{Christlieb}, N.}
\newblock \bibinfo{journal}{\bibinfo{title}{{The Discovery and Analysis of Very
  Metal-Poor Stars in the Galaxy}}}.
\newblock {\emph{\JournalTitle{\araa}}} \textbf{\bibinfo{volume}{43}},
  \bibinfo{pages}{531--580},
  \doiprefix\url{10.1146/annurev.astro.42.053102.134057}
  (\bibinfo{year}{2005}).

\bibitem{Cooke15}
\bibinfo{author}{{Cooke}, R.~J.}, \bibinfo{author}{{Pettini}, M.} \&
  \bibinfo{author}{{Jorgenson}, R.~A.}
\newblock \bibinfo{journal}{\bibinfo{title}{{The Most Metal-poor Damped
  Ly{\ensuremath{\alpha}} Systems: An Insight into Dwarf Galaxies at
  High-redshift}}}.
\newblock {\emph{\JournalTitle{\apj}}} \textbf{\bibinfo{volume}{800}},
  \bibinfo{pages}{12}, \doiprefix\url{10.1088/0004-637X/800/1/12}
  (\bibinfo{year}{2015}).
\newblock \eprint{1406.7003}.

\bibitem{Sodini24}
\bibinfo{author}{{Sodini}, A.} \emph{et~al.}
\newblock \bibinfo{journal}{\bibinfo{title}{{Evidence of Pop III stars'
  chemical signature in neutral gas at z {\ensuremath{\sim}} 6. A study based
  on the E-XQR-30 spectroscopic sample}}}.
\newblock {\emph{\JournalTitle{\aap}}} \textbf{\bibinfo{volume}{687}},
  \bibinfo{pages}{A314}, \doiprefix\url{10.1051/0004-6361/202349062}
  (\bibinfo{year}{2024}).
\newblock \eprint{2404.10722}.

\bibitem{Booth12}
\bibinfo{author}{{Booth}, C.~M.}, \bibinfo{author}{{Schaye}, J.},
  \bibinfo{author}{{Delgado}, J.~D.} \& \bibinfo{author}{{Dalla Vecchia}, C.}
\newblock \bibinfo{journal}{\bibinfo{title}{{The filling factor of
  intergalactic metals at redshift z= 3}}}.
\newblock {\emph{\JournalTitle{\mnras}}} \textbf{\bibinfo{volume}{420}},
  \bibinfo{pages}{1053--1060}, \doiprefix\url{10.1111/j.1365-2966.2011.20047.x}
  (\bibinfo{year}{2012}).
\newblock \eprint{1011.5502}.

\bibitem{Kirihara20}
\bibinfo{author}{{Kirihara}, T.}, \bibinfo{author}{{Hasegawa}, K.},
  \bibinfo{author}{{Umemura}, M.}, \bibinfo{author}{{Mori}, M.} \&
  \bibinfo{author}{{Ishiyama}, T.}
\newblock \bibinfo{journal}{\bibinfo{title}{{Discrimination of heavy elements
  originating from Pop III stars in z = 3 intergalactic medium}}}.
\newblock {\emph{\JournalTitle{\mnras}}} \textbf{\bibinfo{volume}{491}},
  \bibinfo{pages}{4387--4395}, \doiprefix\url{10.1093/mnras/stz3376}
  (\bibinfo{year}{2020}).
\newblock \eprint{1909.07034}.

\bibitem{SAGA}
\bibinfo{author}{{Suda}, T.} \emph{et~al.}
\newblock \bibinfo{journal}{\bibinfo{title}{{Stellar Abundances for the
  Galactic Archeology (SAGA) Database --- Compilation of the Characteristics of
  Known Extremely Metal-Poor Stars}}}.
\newblock {\emph{\JournalTitle{\pasj}}} \textbf{\bibinfo{volume}{60}},
  \bibinfo{pages}{1159}, \doiprefix\url{10.1093/pasj/60.5.1159}
  (\bibinfo{year}{2008}).
\newblock \eprint{0806.3697}.

\bibitem{Omukai05}
\bibinfo{author}{{Omukai}, K.}, \bibinfo{author}{{Tsuribe}, T.},
  \bibinfo{author}{{Schneider}, R.} \& \bibinfo{author}{{Ferrara}, A.}
\newblock \bibinfo{journal}{\bibinfo{title}{{Thermal and Fragmentation
  Properties of Star-forming Clouds in Low-Metallicity Environments}}}.
\newblock {\emph{\JournalTitle{\apj}}} \textbf{\bibinfo{volume}{626}},
  \bibinfo{pages}{627--643}, \doiprefix\url{10.1086/429955}
  (\bibinfo{year}{2005}).
\newblock \eprint{astro-ph/0503010}.

\bibitem{Schneider06}
\bibinfo{author}{{Schneider}, R.}, \bibinfo{author}{{Omukai}, K.},
  \bibinfo{author}{{Inoue}, A.~K.} \& \bibinfo{author}{{Ferrara}, A.}
\newblock \bibinfo{journal}{\bibinfo{title}{{Fragmentation of star-forming
  clouds enriched with the first dust}}}.
\newblock {\emph{\JournalTitle{\mnras}}} \textbf{\bibinfo{volume}{369}},
  \bibinfo{pages}{1437--1444}, \doiprefix\url{10.1111/j.1365-2966.2006.10391.x}
  (\bibinfo{year}{2006}).
\newblock \eprint{astro-ph/0603766}.

\bibitem{Chon21}
\bibinfo{author}{{Chon}, S.}, \bibinfo{author}{{Omukai}, K.} \&
  \bibinfo{author}{{Schneider}, R.}
\newblock \bibinfo{journal}{\bibinfo{title}{{Transition of the initial mass
  function in the metal-poor environments}}}.
\newblock {\emph{\JournalTitle{\mnras}}} \textbf{\bibinfo{volume}{508}},
  \bibinfo{pages}{4175--4192}, \doiprefix\url{10.1093/mnras/stab2497}
  (\bibinfo{year}{2021}).
\newblock \eprint{2103.04997}.

\bibitem{BSmith24}
\bibinfo{author}{{Smith}, B.~D.} \emph{et~al.}
\newblock \bibinfo{journal}{\bibinfo{title}{{Why does the Milky Way have a
  metallicity floor?}}}
\newblock {\emph{\JournalTitle{\mnras}}} \textbf{\bibinfo{volume}{532}},
  \bibinfo{pages}{3797--3807}, \doiprefix\url{10.1093/mnras/stae1725}
  (\bibinfo{year}{2024}).
\newblock \eprint{2406.08199}.

\bibitem{Chiaki22}
\bibinfo{author}{{Chiaki}, G.} \& \bibinfo{author}{{Yoshida}, N.}
\newblock \bibinfo{journal}{\bibinfo{title}{{Disc fragmentation and oligarchic
  growth of protostellar systems in low-metallicity gas clouds}}}.
\newblock {\emph{\JournalTitle{\mnras}}} \textbf{\bibinfo{volume}{510}},
  \bibinfo{pages}{5199--5219}, \doiprefix\url{10.1093/mnras/stab2799}
  (\bibinfo{year}{2022}).
\newblock \eprint{2008.06107}.

\bibitem{Bosman22}
\bibinfo{author}{{Bosman}, S. E.~I.} \emph{et~al.}
\newblock \bibinfo{journal}{\bibinfo{title}{{Hydrogen reionization ends by z =
  5.3: Lyman-{\ensuremath{\alpha}} optical depth measured by the XQR-30
  sample}}}.
\newblock {\emph{\JournalTitle{\mnras}}} \textbf{\bibinfo{volume}{514}},
  \bibinfo{pages}{55--76}, \doiprefix\url{10.1093/mnras/stac1046}
  (\bibinfo{year}{2022}).
\newblock \eprint{2108.03699}.

\bibitem{Limongi18}
\bibinfo{author}{{Limongi}, M.} \& \bibinfo{author}{{Chieffi}, A.}
\newblock \bibinfo{journal}{\bibinfo{title}{{Presupernova Evolution and
  Explosive Nucleosynthesis of Rotating Massive Stars in the Metallicity Range
  -3 {\ensuremath{\leq}} [Fe/H] {\ensuremath{\leq}} 0}}}.
\newblock {\emph{\JournalTitle{\apjs}}} \textbf{\bibinfo{volume}{237}},
  \bibinfo{pages}{13}, \doiprefix\url{10.3847/1538-4365/aacb24}
  (\bibinfo{year}{2018}).
\newblock \eprint{1805.09640}.

\bibitem{Heger10}
\bibinfo{author}{{Heger}, A.} \& \bibinfo{author}{{Woosley}, S.~E.}
\newblock \bibinfo{journal}{\bibinfo{title}{{Nucleosynthesis and Evolution of
  Massive Metal-free Stars}}}.
\newblock {\emph{\JournalTitle{\apj}}} \textbf{\bibinfo{volume}{724}},
  \bibinfo{pages}{341--373}, \doiprefix\url{10.1088/0004-637X/724/1/341}
  (\bibinfo{year}{2010}).
\newblock \eprint{0803.3161}.

\bibitem{Welsh24}
\bibinfo{author}{{Welsh}, L.}, \bibinfo{author}{{Cooke}, R.},
  \bibinfo{author}{{Fumagalli}, M.}, \bibinfo{author}{{Pettini}, M.} \&
  \bibinfo{author}{{Rudie}, G.~C.}
\newblock \bibinfo{journal}{\bibinfo{title}{{A survey of extremely metal-poor
  gas at cosmic noon: Evidence of elevated [O/Fe]}}}.
\newblock {\emph{\JournalTitle{\aap}}} \textbf{\bibinfo{volume}{691}},
  \bibinfo{pages}{A285}, \doiprefix\url{10.1051/0004-6361/202451147}
  (\bibinfo{year}{2024}).
\newblock \eprint{2409.07525}.

\bibitem{Trenti09}
\bibinfo{author}{{Trenti}, M.}, \bibinfo{author}{{Stiavelli}, M.} \&
  \bibinfo{author}{{Shull}, J.~M.}
\newblock \bibinfo{journal}{\bibinfo{title}{{Metal-free Gas Supply at the Edge
  of Reionization: Late-epoch Population III Star Formation}}}.
\newblock {\emph{\JournalTitle{\apj}}} \textbf{\bibinfo{volume}{700}},
  \bibinfo{pages}{1672--1679}, \doiprefix\url{10.1088/0004-637X/700/2/1672}
  (\bibinfo{year}{2009}).
\newblock \eprint{0905.4504}.

\bibitem{Liu20}
\bibinfo{author}{{Liu}, B.} \& \bibinfo{author}{{Bromm}, V.}
\newblock \bibinfo{journal}{\bibinfo{title}{{When did Population III star
  formation end?}}}
\newblock {\emph{\JournalTitle{\mnras}}} \textbf{\bibinfo{volume}{497}},
  \bibinfo{pages}{2839--2854}, \doiprefix\url{10.1093/mnras/staa2143}
  (\bibinfo{year}{2020}).
\newblock \eprint{2006.15260}.

\bibitem{vanVeenen25}
\bibinfo{author}{{van Veenen}, L.}, \bibinfo{author}{{Sharda}, P.},
  \bibinfo{author}{{Viti}, S.} \& \bibinfo{author}{{Menon}, S.~H.}
\newblock \bibinfo{journal}{\bibinfo{title}{{Radiation magnetohydrodynamics
  simulations of Population III star formation during the epoch of
  reionization}}}.
\newblock {\emph{\JournalTitle{\aap}}} \textbf{\bibinfo{volume}{708}},
  \bibinfo{pages}{A61}, \doiprefix\url{10.1051/0004-6361/202558114}
  (\bibinfo{year}{2026}).
\newblock \eprint{2511.11314}.

\bibitem{Caffau11}
\bibinfo{author}{{Caffau}, E.} \emph{et~al.}
\newblock \bibinfo{journal}{\bibinfo{title}{{An extremely primitive star in the
  Galactic halo}}}.
\newblock {\emph{\JournalTitle{\nat}}} \textbf{\bibinfo{volume}{477}},
  \bibinfo{pages}{67--69}, \doiprefix\url{10.1038/nature10377}
  (\bibinfo{year}{2011}).
\newblock \eprint{1203.2612}.

\bibitem{Lagae23}
\bibinfo{author}{{Lagae}, C.} \emph{et~al.}
\newblock \bibinfo{journal}{\bibinfo{title}{{Raising the observed metallicity
  floor with a 3D non-LTE analysis of SDSS J102915.14+172927.9}}}.
\newblock {\emph{\JournalTitle{\aap}}} \textbf{\bibinfo{volume}{672}},
  \bibinfo{pages}{A90}, \doiprefix\url{10.1051/0004-6361/202245786}
  (\bibinfo{year}{2023}).
\newblock \eprint{2303.01374}.

\bibitem{Caffau24}
\bibinfo{author}{{Caffau}, E.} \emph{et~al.}
\newblock \bibinfo{journal}{\bibinfo{title}{{SDSS J102915.14+172927.9:
  Revisiting the chemical pattern}}}.
\newblock {\emph{\JournalTitle{\aap}}} \textbf{\bibinfo{volume}{691}},
  \bibinfo{pages}{A245}, \doiprefix\url{10.1051/0004-6361/202452079}
  (\bibinfo{year}{2024}).
\newblock \eprint{2411.13096}.

\bibitem{Limberg24}
\bibinfo{author}{{Limberg}, G.} \emph{et~al.}
\newblock \bibinfo{journal}{\bibinfo{title}{{Discovery of an [Fe/H]
  {\ensuremath{\sim}} ‑4.8 Star in Gaia XP Spectra}}}.
\newblock {\emph{\JournalTitle{\apjl}}} \textbf{\bibinfo{volume}{989}},
  \bibinfo{pages}{L18}, \doiprefix\url{10.3847/2041-8213/adf196}
  (\bibinfo{year}{2025}).
\newblock \eprint{2508.00067}.

\bibitem{Ji25}
\bibinfo{author}{{Ji}, A.~P.} \emph{et~al.}
\newblock \bibinfo{journal}{\bibinfo{title}{{A nearly pristine star from the
  Large Magellanic Cloud}}}.
\newblock {\emph{\JournalTitle{Nature Astronomy}}}
  \doiprefix\url{10.1038/s41550-026-02816-7} (\bibinfo{year}{2026}).
\newblock \eprint{2509.21643}.

\bibitem{Saccardi23}
\bibinfo{author}{{Saccardi}, A.} \emph{et~al.}
\newblock \bibinfo{journal}{\bibinfo{title}{{Evidence of First Stars-enriched
  Gas in High-redshift Absorbers}}}.
\newblock {\emph{\JournalTitle{\apj}}} \textbf{\bibinfo{volume}{948}},
  \bibinfo{pages}{35}, \doiprefix\url{10.3847/1538-4357/acc39f}
  (\bibinfo{year}{2023}).
\newblock \eprint{2305.02346}.

\bibitem{Rafelski12}
\bibinfo{author}{{Rafelski}, M.}, \bibinfo{author}{{Wolfe}, A.~M.},
  \bibinfo{author}{{Prochaska}, J.~X.}, \bibinfo{author}{{Neeleman}, M.} \&
  \bibinfo{author}{{Mendez}, A.~J.}
\newblock \bibinfo{journal}{\bibinfo{title}{{Metallicity Evolution of Damped
  Ly{\ensuremath{\alpha}} Systems Out to z \raisebox{-0.5ex}\textasciitilde
  5}}}.
\newblock {\emph{\JournalTitle{\apj}}} \textbf{\bibinfo{volume}{755}},
  \bibinfo{pages}{89}, \doiprefix\url{10.1088/0004-637X/755/2/89}
  (\bibinfo{year}{2012}).
\newblock \eprint{1205.5047}.

\bibitem{Rafelski14}
\bibinfo{author}{{Rafelski}, M.}, \bibinfo{author}{{Neeleman}, M.},
  \bibinfo{author}{{Fumagalli}, M.}, \bibinfo{author}{{Wolfe}, A.~M.} \&
  \bibinfo{author}{{Prochaska}, J.~X.}
\newblock \bibinfo{journal}{\bibinfo{title}{{The Rapid Decline in Metallicity
  of Damped Ly{\ensuremath{\alpha}} Systems at z
  \raisebox{-0.5ex}\textasciitilde 5}}}.
\newblock {\emph{\JournalTitle{\apjl}}} \textbf{\bibinfo{volume}{782}},
  \bibinfo{pages}{L29}, \doiprefix\url{10.1088/2041-8205/782/2/L29}
  (\bibinfo{year}{2014}).
\newblock \eprint{1310.6042}.

\bibitem{D'Odorico23}
\bibinfo{author}{{D'Odorico}, V.} \emph{et~al.}
\newblock \bibinfo{journal}{\bibinfo{title}{{XQR-30: The ultimate XSHOOTER
  quasar sample at the reionization epoch}}}.
\newblock {\emph{\JournalTitle{\mnras}}} \textbf{\bibinfo{volume}{523}},
  \bibinfo{pages}{1399--1420}, \doiprefix\url{10.1093/mnras/stad1468}
  (\bibinfo{year}{2023}).
\newblock \eprint{2305.05053}.

\bibitem{Simcoe12}
\bibinfo{author}{{Simcoe}, R.~A.} \emph{et~al.}
\newblock \bibinfo{journal}{\bibinfo{title}{{Extremely metal-poor gas at a
  redshift of 7}}}.
\newblock {\emph{\JournalTitle{\nat}}} \textbf{\bibinfo{volume}{492}},
  \bibinfo{pages}{79--82}, \doiprefix\url{10.1038/nature11612}
  (\bibinfo{year}{2012}).
\newblock \eprint{1212.0548}.

\bibitem{Davies25-PDLA}
\bibinfo{author}{{Davies}, F.~B.}, \bibinfo{author}{{Ba{\~n}ados}, E.},
  \bibinfo{author}{{Hennawi}, J.~F.} \& \bibinfo{author}{{Bosman}, S. E.~I.}
\newblock \bibinfo{journal}{\bibinfo{title}{{Possible Contamination of the
  Intergalactic Medium Damping Wing in ULAS J1342+0928 by Proximate Damped
  Ly{\ensuremath{\alpha}} Absorption}}}.
\newblock {\emph{\JournalTitle{\apjl}}} \textbf{\bibinfo{volume}{989}},
  \bibinfo{pages}{L27}, \doiprefix\url{10.3847/2041-8213/ade7fb}
  (\bibinfo{year}{2025}).
\newblock \eprint{2312.06747}.

\bibitem{D'Odorico24}
\bibinfo{author}{{D'Odorico}, V.} \emph{et~al.}
\newblock \bibinfo{journal}{\bibinfo{title}{{Galaxy formation and symbiotic
  evolution with the inter-galactic medium in the age of ELT-ANDES}}}.
\newblock {\emph{\JournalTitle{Experimental Astronomy}}}
  \textbf{\bibinfo{volume}{58}}, \bibinfo{pages}{21},
  \doiprefix\url{10.1007/s10686-024-09967-3} (\bibinfo{year}{2024}).
\newblock \eprint{2311.16803}.

\bibitem{Euclid-quasars}
\bibinfo{author}{{Euclid Collaboration}} \emph{et~al.}
\newblock \bibinfo{journal}{\bibinfo{title}{{Euclid preparation. V. Predicted
  yield of redshift 7 < z < 9 quasars from the wide survey}}}.
\newblock {\emph{\JournalTitle{\aap}}} \textbf{\bibinfo{volume}{631}},
  \bibinfo{pages}{A85}, \doiprefix\url{10.1051/0004-6361/201936427}
  (\bibinfo{year}{2019}).
\newblock \eprint{1908.04310}.


\end{thebibliography}

\begin{thebibliography}{100}
\urlstyle{rm}
\expandafter\ifx\csname url\endcsname\relax
  \def\url#1{\texttt{#1}}\fi
\expandafter\ifx\csname urlprefix\endcsname\relax\def\urlprefix{URL }\fi
\expandafter\ifx\csname doiprefix\endcsname\relax\def\doiprefix{DOI: }\fi
\providecommand{\bibinfo}[2]{#2}
\providecommand{\eprint}[2][]{\url{#2}}


\setcounter{enumiv}{50}


\bibitem{Planck20}
\bibinfo{author}{{Planck Collaboration}} \emph{et~al.}
\newblock \bibinfo{journal}{\bibinfo{title}{{Planck 2018 results. VI.
  Cosmological parameters}}}.
\newblock {\emph{\JournalTitle{\aap}}} \textbf{\bibinfo{volume}{641}},
  \bibinfo{pages}{A6}, \doiprefix\url{10.1051/0004-6361/201833910}
  (\bibinfo{year}{2020}).
\newblock \eprint{1807.06209}.

\bibitem{Asplund09}
\bibinfo{author}{{Asplund}, M.}, \bibinfo{author}{{Grevesse}, N.},
  \bibinfo{author}{{Sauval}, A.~J.} \& \bibinfo{author}{{Scott}, P.}
\newblock \bibinfo{journal}{\bibinfo{title}{{The Chemical Composition of the
  Sun}}}.
\newblock {\emph{\JournalTitle{\araa}}} \textbf{\bibinfo{volume}{47}},
  \bibinfo{pages}{481--522},
  \doiprefix\url{10.1146/annurev.astro.46.060407.145222}
  (\bibinfo{year}{2009}).
\newblock \eprint{0909.0948}.

\bibitem{Magg22}
\bibinfo{author}{{Magg}, E.} \emph{et~al.}
\newblock \bibinfo{journal}{\bibinfo{title}{{Observational constraints on the
  origin of the elements. IV. Standard composition of the Sun}}}.
\newblock {\emph{\JournalTitle{\aap}}} \textbf{\bibinfo{volume}{661}},
  \bibinfo{pages}{A140}, \doiprefix\url{10.1051/0004-6361/202142971}
  (\bibinfo{year}{2022}).
\newblock \eprint{2203.02255}.

\bibitem{GS98}
\bibinfo{author}{{Grevesse}, N.} \& \bibinfo{author}{{Sauval}, A.~J.}
\newblock \bibinfo{journal}{\bibinfo{title}{{Standard Solar Composition}}}.
\newblock {\emph{\JournalTitle{\ssr}}} \textbf{\bibinfo{volume}{85}},
  \bibinfo{pages}{161--174}, \doiprefix\url{10.1023/A:1005161325181}
  (\bibinfo{year}{1998}).

\bibitem{Onken22}
\bibinfo{author}{{Onken}, C.~A.} \emph{et~al.}
\newblock \bibinfo{journal}{\bibinfo{title}{{Ultraluminous high-redshift
  quasars from SkyMapper - II. New quasars and the bright end of the luminosity
  function}}}.
\newblock {\emph{\JournalTitle{\mnras}}} \textbf{\bibinfo{volume}{511}},
  \bibinfo{pages}{572--594}, \doiprefix\url{10.1093/mnras/stac051}
  (\bibinfo{year}{2022}).
\newblock \eprint{2105.12215}.

\bibitem{Lai24}
\bibinfo{author}{{Lai}, S.}, \bibinfo{author}{{Onken}, C.~A.},
  \bibinfo{author}{{Wolf}, C.}, \bibinfo{author}{{Bian}, F.} \&
  \bibinfo{author}{{Fan}, X.}
\newblock \bibinfo{journal}{\bibinfo{title}{{XQz5: a new ultraluminous z
  {\ensuremath{\sim}} 5 quasar legacy sample}}}.
\newblock {\emph{\JournalTitle{\mnras}}} \textbf{\bibinfo{volume}{527}},
  \bibinfo{pages}{3912--3931}, \doiprefix\url{10.1093/mnras/stad3474}
  (\bibinfo{year}{2024}).
\newblock \eprint{2311.05098}.

\bibitem{DRAGONS1}
\bibinfo{author}{{Labrie}, K.} \emph{et~al.}
\newblock \bibinfo{journal}{\bibinfo{title}{{DRAGONS-A Quick Overview}}}.
\newblock {\emph{\JournalTitle{Research Notes of the American Astronomical
  Society}}} \textbf{\bibinfo{volume}{7}}, \bibinfo{pages}{214},
  \doiprefix\url{10.3847/2515-5172/ad0044} (\bibinfo{year}{2023}).
\newblock \eprint{2310.03048}.

\bibitem{DRAGONS2}
\bibinfo{author}{{Simpson}, C.} \emph{et~al.}
\newblock \bibinfo{title}{{DRAGONS}}, \doiprefix\url{10.5281/zenodo.4025470}
  (\bibinfo{year}{2025}).

\bibitem{Pypeit-official}
\bibinfo{author}{{Prochaska}, J.} \emph{et~al.}
\newblock \bibinfo{journal}{\bibinfo{title}{{PypeIt: The Python Spectroscopic
  Data Reduction Pipeline}}}.
\newblock {\emph{\JournalTitle{The Journal of Open Source Software}}}
  \textbf{\bibinfo{volume}{5}}, \bibinfo{pages}{2308},
  \doiprefix\url{10.21105/joss.02308} (\bibinfo{year}{2020}).
\newblock \eprint{2005.06505}.

\bibitem{LACosmic}
\bibinfo{author}{{van Dokkum}, P.~G.}
\newblock \bibinfo{journal}{\bibinfo{title}{{Cosmic-Ray Rejection by Laplacian
  Edge Detection}}}.
\newblock {\emph{\JournalTitle{\pasp}}} \textbf{\bibinfo{volume}{113}},
  \bibinfo{pages}{1420--1427}, \doiprefix\url{10.1086/323894}
  (\bibinfo{year}{2001}).
\newblock \eprint{astro-ph/0108003}.

\bibitem{Horne86}
\bibinfo{author}{{Horne}, K.}
\newblock \bibinfo{journal}{\bibinfo{title}{{An optimal extraction algorithm
  for CCD spectroscopy}}}.
\newblock {\emph{\JournalTitle{\pasp}}} \textbf{\bibinfo{volume}{98}},
  \bibinfo{pages}{609--617}, \doiprefix\url{10.1086/131801}
  (\bibinfo{year}{1986}).

\bibitem{Kelson03}
\bibinfo{author}{{Kelson}, D.~D.}
\newblock \bibinfo{journal}{\bibinfo{title}{{Optimal Techniques in
  Two-dimensional Spectroscopy: Background Subtraction for the 21st Century}}}.
\newblock {\emph{\JournalTitle{\pasp}}} \textbf{\bibinfo{volume}{115}},
  \bibinfo{pages}{688--699}, \doiprefix\url{10.1086/375502}
  (\bibinfo{year}{2003}).
\newblock \eprint{astro-ph/0303507}.

\bibitem{LBLRTM}
\bibinfo{author}{{Clough}, S.~A.} \emph{et~al.}
\newblock \bibinfo{journal}{\bibinfo{title}{{Atmospheric radiative transfer
  modeling: a summary of the AER codes}}}.
\newblock {\emph{\JournalTitle{\jqsrt}}} \textbf{\bibinfo{volume}{91}},
  \bibinfo{pages}{233--244}, \doiprefix\url{10.1016/j.jqsrt.2004.05.058}
  (\bibinfo{year}{2005}).

\bibitem{TelFit}
\bibinfo{author}{{Gullikson}, K.}, \bibinfo{author}{{Dodson-Robinson}, S.} \&
  \bibinfo{author}{{Kraus}, A.}
\newblock \bibinfo{journal}{\bibinfo{title}{{Correcting for Telluric
  Absorption: Methods, Case Studies, and Release of the TelFit Code}}}.
\newblock {\emph{\JournalTitle{\aj}}} \textbf{\bibinfo{volume}{148}},
  \bibinfo{pages}{53}, \doiprefix\url{10.1088/0004-6256/148/3/53}
  (\bibinfo{year}{2014}).
\newblock \eprint{1406.6059}.

\bibitem{Greig17}
\bibinfo{author}{{Greig}, B.}, \bibinfo{author}{{Mesinger}, A.},
  \bibinfo{author}{{Haiman}, Z.} \& \bibinfo{author}{{Simcoe}, R.~A.}
\newblock \bibinfo{journal}{\bibinfo{title}{{Are we witnessing the epoch of
  reionisation at z = 7.1 from the spectrum of J1120+0641?}}}
\newblock {\emph{\JournalTitle{\mnras}}} \textbf{\bibinfo{volume}{466}},
  \bibinfo{pages}{4239--4249}, \doiprefix\url{10.1093/mnras/stw3351}
  (\bibinfo{year}{2017}).
\newblock \eprint{1606.00441}.

\bibitem{Bolton17}
\bibinfo{author}{{Bolton}, J.~S.} \emph{et~al.}
\newblock \bibinfo{journal}{\bibinfo{title}{{The Sherwood simulation suite:
  overview and data comparisons with the Lyman {$\alpha$} forest at redshifts
  $2 \leq z \leq 5$}}}.
\newblock {\emph{\JournalTitle{\mnras}}} \textbf{\bibinfo{volume}{464}},
  \bibinfo{pages}{897--914}, \doiprefix\url{10.1093/mnras/stw2397}
  (\bibinfo{year}{2017}).
\newblock \eprint{1605.03462}.

\bibitem{Viel13}
\bibinfo{author}{{Viel}, M.}, \bibinfo{author}{{Becker}, G.~D.},
  \bibinfo{author}{{Bolton}, J.~S.} \& \bibinfo{author}{{Haehnelt}, M.~G.}
\newblock \bibinfo{journal}{\bibinfo{title}{{Warm dark matter as a solution to
  the small scale crisis: New constraints from high redshift Lyman-{$\alpha$}
  forest data}}}.
\newblock {\emph{\JournalTitle{\prd}}} \textbf{\bibinfo{volume}{88}},
  \bibinfo{pages}{043502}, \doiprefix\url{10.1103/PhysRevD.88.043502}
  (\bibinfo{year}{2013}).
\newblock \eprint{1306.2314}.

\bibitem{Oyarzun25}
\bibinfo{author}{{Oyarz{\'u}n}, G.~A.} \emph{et~al.}
\newblock \bibinfo{journal}{\bibinfo{title}{{The Qz5 Survey. I. How the H I
  Mass Density of the Universe Evolves with Cosmic Time}}}.
\newblock {\emph{\JournalTitle{\apj}}} \textbf{\bibinfo{volume}{983}},
  \bibinfo{pages}{10}, \doiprefix\url{10.3847/1538-4357/adb430}
  (\bibinfo{year}{2025}).
\newblock \eprint{2502.05261}.

\bibitem{Bajtlik88}
\bibinfo{author}{{Bajtlik}, S.}, \bibinfo{author}{{Duncan}, R.~C.} \&
  \bibinfo{author}{{Ostriker}, J.~P.}
\newblock \bibinfo{journal}{\bibinfo{title}{{Quasar Ionization of Lyman-Alpha
  Clouds: The Proximity Effect, a Probe of the Ultraviolet Background at High
  Redshift}}}.
\newblock {\emph{\JournalTitle{\apj}}} \textbf{\bibinfo{volume}{327}},
  \bibinfo{pages}{570}, \doiprefix\url{10.1086/166217} (\bibinfo{year}{1988}).

\bibitem{Davies20-prox}
\bibinfo{author}{{Davies}, F.~B.}, \bibinfo{author}{{Hennawi}, J.~F.} \&
  \bibinfo{author}{{Eilers}, A.-C.}
\newblock \bibinfo{journal}{\bibinfo{title}{{Time-dependent behaviour of quasar
  proximity zones at z {\ensuremath{\sim}} 6}}}.
\newblock {\emph{\JournalTitle{\mnras}}} \textbf{\bibinfo{volume}{493}},
  \bibinfo{pages}{1330--1343}, \doiprefix\url{10.1093/mnras/stz3303}
  (\bibinfo{year}{2020}).
\newblock \eprint{1903.12346}.

\bibitem{Lusso15}
\bibinfo{author}{{Lusso}, E.} \emph{et~al.}
\newblock \bibinfo{journal}{\bibinfo{title}{{The first ultraviolet
  quasar-stacked spectrum at z {$\sim$} 2.4 from WFC3}}}.
\newblock {\emph{\JournalTitle{\mnras}}} \textbf{\bibinfo{volume}{449}},
  \bibinfo{pages}{4204--4220}, \doiprefix\url{10.1093/mnras/stv516}
  (\bibinfo{year}{2015}).
\newblock \eprint{1503.02075}.

\bibitem{Gaikwad23}
\bibinfo{author}{{Gaikwad}, P.} \emph{et~al.}
\newblock \bibinfo{journal}{\bibinfo{title}{{Measuring the photoionization
  rate, neutral fraction, and mean free path of H I ionizing photons at 4.9
  {\ensuremath{\leq}} z {\ensuremath{\leq}} 6.0 from a large sample of XShooter
  and ESI spectra}}}.
\newblock {\emph{\JournalTitle{\mnras}}} \textbf{\bibinfo{volume}{525}},
  \bibinfo{pages}{4093--4120}, \doiprefix\url{10.1093/mnras/stad2566}
  (\bibinfo{year}{2023}).
\newblock \eprint{2304.02038}.

\bibitem{Davies24}
\bibinfo{author}{{Davies}, F.~B.} \emph{et~al.}
\newblock \bibinfo{journal}{\bibinfo{title}{{Constraints on the Evolution of
  the Ionizing Background and Ionizing Photon Mean Free Path at the End of
  Reionization}}}.
\newblock {\emph{\JournalTitle{\apj}}} \textbf{\bibinfo{volume}{965}},
  \bibinfo{pages}{134}, \doiprefix\url{10.3847/1538-4357/ad1d5d}
  (\bibinfo{year}{2024}).
\newblock \eprint{2312.08464}.

\bibitem{Maity26}
\bibinfo{author}{{Maity}, B.}, \bibinfo{author}{{Davies}, F.} \&
  \bibinfo{author}{{Gaikwad}, P.}
\newblock \bibinfo{journal}{\bibinfo{title}{{Efficient modeling of
  Lyman-{\ensuremath{\alpha}} opacity fluctuations during the late reionization
  epoch}}}.
\newblock {\emph{\JournalTitle{\aap}}} \textbf{\bibinfo{volume}{706}},
  \bibinfo{pages}{A36}, \doiprefix\url{10.1051/0004-6361/202555515}
  (\bibinfo{year}{2026}).
\newblock \eprint{2505.09369}.

\bibitem{Jorgenson09}
\bibinfo{author}{{Jorgenson}, R.~A.}, \bibinfo{author}{{Wolfe}, A.~M.},
  \bibinfo{author}{{Prochaska}, J.~X.} \& \bibinfo{author}{{Carswell}, R.~F.}
\newblock \bibinfo{journal}{\bibinfo{title}{{Direct Evidence of Cold Gas in DLA
  0812+32B}}}.
\newblock {\emph{\JournalTitle{\apj}}} \textbf{\bibinfo{volume}{704}},
  \bibinfo{pages}{247--254}, \doiprefix\url{10.1088/0004-637X/704/1/247}
  (\bibinfo{year}{2009}).
\newblock \eprint{0908.3485}.

\bibitem{Fumagalli11}
\bibinfo{author}{{Fumagalli}, M.}, \bibinfo{author}{{O'Meara}, J.~M.} \&
  \bibinfo{author}{{Prochaska}, J.~X.}
\newblock \bibinfo{journal}{\bibinfo{title}{{Detection of Pristine Gas Two
  Billion Years After the Big Bang}}}.
\newblock {\emph{\JournalTitle{Science}}} \textbf{\bibinfo{volume}{334}},
  \bibinfo{pages}{1245}, \doiprefix\url{10.1126/science.1213581}
  (\bibinfo{year}{2011}).
\newblock \eprint{1111.2334}.

\bibitem{Bergemann25}
\bibinfo{author}{{Bergemann}, M.} \& \bibinfo{author}{{Hoppe}, R.}
\newblock \bibinfo{journal}{\bibinfo{title}{{3D Non-LTE radiation transfer:
  theory and applications to stars, exoplanets, and kilonovae}}}.
\newblock {\emph{\JournalTitle{arXiv e-prints}}}
  \bibinfo{pages}{arXiv:2511.04254}, \doiprefix\url{10.48550/arXiv.2511.04254}
  (\bibinfo{year}{2025}).
\newblock \eprint{2511.04254}.

\bibitem{Wolfe94}
\bibinfo{author}{{Wolfe}, A.~M.} \emph{et~al.}
\newblock \bibinfo{journal}{\bibinfo{title}{{Metal Abundances and Kinematics of
  a High-Redshift Galaxy Obtained with the Keck Telescope}}}.
\newblock {\emph{\JournalTitle{\apjl}}} \textbf{\bibinfo{volume}{435}},
  \bibinfo{pages}{L101}, \doiprefix\url{10.1086/187604} (\bibinfo{year}{1994}).

\bibitem{Lu96}
\bibinfo{author}{{Lu}, L.}, \bibinfo{author}{{Sargent}, W. L.~W.},
  \bibinfo{author}{{Barlow}, T.~A.}, \bibinfo{author}{{Churchill}, C.~W.} \&
  \bibinfo{author}{{Vogt}, S.~S.}
\newblock \bibinfo{journal}{\bibinfo{title}{{Abundances at High Redshifts: The
  Chemical Enrichment History of Damped Ly alpha Galaxies}}}.
\newblock {\emph{\JournalTitle{\apjs}}} \textbf{\bibinfo{volume}{107}},
  \bibinfo{pages}{475}, \doiprefix\url{10.1086/192373} (\bibinfo{year}{1996}).
\newblock \eprint{astro-ph/9606044}.

\bibitem{Prochaska96}
\bibinfo{author}{{Prochaska}, J.~X.} \& \bibinfo{author}{{Wolfe}, A.~M.}
\newblock \bibinfo{journal}{\bibinfo{title}{{A Keck HIRES Investigation of the
  Metal Abundances and Kinematics of the Z = 2.46 Damped LY alpha System toward
  Q0201+365}}}.
\newblock {\emph{\JournalTitle{\apj}}} \textbf{\bibinfo{volume}{470}},
  \bibinfo{pages}{403}, \doiprefix\url{10.1086/177875} (\bibinfo{year}{1996}).
\newblock \eprint{astro-ph/9604042}.

\bibitem{Prochaska97a}
\bibinfo{author}{{Prochaska}, J.~X.} \& \bibinfo{author}{{Wolfe}, M.}
\newblock \bibinfo{journal}{\bibinfo{title}{{A Keck HIRES Investigation of the
  Metal Abundances and Kinematics of Three Damped Ly{\ensuremath{\alpha}}
  Systems toward Q2206-199}}}.
\newblock {\emph{\JournalTitle{\apj}}} \textbf{\bibinfo{volume}{474}},
  \bibinfo{pages}{140--178}, \doiprefix\url{10.1086/303457}
  (\bibinfo{year}{1997}).
\newblock \eprint{astro-ph/9605021}.

\bibitem{Lu98}
\bibinfo{author}{{Lu}, L.}, \bibinfo{author}{{Sargent}, W. L.~W.} \&
  \bibinfo{author}{{Barlow}, T.~A.}
\newblock \bibinfo{journal}{\bibinfo{title}{{The N/Si Abundance Ratio in 15
  Damped Lyalpha Galaxies: Implications for the Origin of Nitrogen}}}.
\newblock {\emph{\JournalTitle{\aj}}} \textbf{\bibinfo{volume}{115}},
  \bibinfo{pages}{55--61}, \doiprefix\url{10.1086/300180}
  (\bibinfo{year}{1998}).
\newblock \eprint{astro-ph/9710266}.

\bibitem{Lopez99}
\bibinfo{author}{{Lopez}, S.}, \bibinfo{author}{{Reimers}, D.},
  \bibinfo{author}{{Rauch}, M.}, \bibinfo{author}{{Sargent}, W. L.~W.} \&
  \bibinfo{author}{{Smette}, A.}
\newblock \bibinfo{journal}{\bibinfo{title}{{First Comparison of Ionization and
  Metallicity in Two Lines of Sight toward HE 1104-1805 AB at z=1.66}}}.
\newblock {\emph{\JournalTitle{\apj}}} \textbf{\bibinfo{volume}{513}},
  \bibinfo{pages}{598--618}, \doiprefix\url{10.1086/306879}
  (\bibinfo{year}{1999}).
\newblock \eprint{astro-ph/9806143}.

\bibitem{Pettini99}
\bibinfo{author}{{Pettini}, M.}, \bibinfo{author}{{Ellison}, S.~L.},
  \bibinfo{author}{{Steidel}, C.~C.} \& \bibinfo{author}{{Bowen}, D.~V.}
\newblock \bibinfo{journal}{\bibinfo{title}{{Metal Abundances at z<1.5: Fresh
  Clues to the Chemical Enrichment History of Damped Ly{\ensuremath{\alpha}}
  Systems}}}.
\newblock {\emph{\JournalTitle{\apj}}} \textbf{\bibinfo{volume}{510}},
  \bibinfo{pages}{576--589}, \doiprefix\url{10.1086/306635}
  (\bibinfo{year}{1999}).
\newblock \eprint{astro-ph/9808017}.

\bibitem{Prochaska99}
\bibinfo{author}{{Prochaska}, J.~X.} \& \bibinfo{author}{{Wolfe}, A.~M.}
\newblock \bibinfo{journal}{\bibinfo{title}{{Chemical Abundances of the Damped
  Ly{\ensuremath{\alpha}} Systems at z>1.5}}}.
\newblock {\emph{\JournalTitle{\apjs}}} \textbf{\bibinfo{volume}{121}},
  \bibinfo{pages}{369--415}, \doiprefix\url{10.1086/313200}
  (\bibinfo{year}{1999}).
\newblock \eprint{astro-ph/9810381}.

\bibitem{Molaro00}
\bibinfo{author}{{Molaro}, P.} \emph{et~al.}
\newblock \bibinfo{journal}{\bibinfo{title}{{UVES Observations of QSO
  0000-2620: Oxygen and Zinc Abundances in the Damped Ly{\ensuremath{\alpha}}
  Galaxy at Z$_{abs}$=3.3901}}}.
\newblock {\emph{\JournalTitle{\apj}}} \textbf{\bibinfo{volume}{541}},
  \bibinfo{pages}{54--60}, \doiprefix\url{10.1086/309439}
  (\bibinfo{year}{2000}).
\newblock \eprint{astro-ph/0005098}.

\bibitem{Petitjean00}
\bibinfo{author}{{Petitjean}, P.}, \bibinfo{author}{{Srianand}, R.} \&
  \bibinfo{author}{{Ledoux}, C.}
\newblock \bibinfo{journal}{\bibinfo{title}{{Molecular hydrogen and the nature
  of damped Lyman-alpha systems}}}.
\newblock {\emph{\JournalTitle{\aap}}} \textbf{\bibinfo{volume}{364}},
  \bibinfo{pages}{L26--L30}, \doiprefix\url{10.48550/arXiv.astro-ph/0011437}
  (\bibinfo{year}{2000}).
\newblock \eprint{astro-ph/0011437}.

\bibitem{Pettini00}
\bibinfo{author}{{Pettini}, M.}, \bibinfo{author}{{Ellison}, S.~L.},
  \bibinfo{author}{{Steidel}, C.~C.}, \bibinfo{author}{{Shapley}, A.~E.} \&
  \bibinfo{author}{{Bowen}, D.~V.}
\newblock \bibinfo{journal}{\bibinfo{title}{{Si and Mn Abundances in Damped
  Ly{\ensuremath{\alpha}} Systems with Low Dust Content}}}.
\newblock {\emph{\JournalTitle{\apj}}} \textbf{\bibinfo{volume}{532}},
  \bibinfo{pages}{65--76}, \doiprefix\url{10.1086/308562}
  (\bibinfo{year}{2000}).
\newblock \eprint{astro-ph/9910131}.

\bibitem{Prochaska00}
\bibinfo{author}{{Prochaska}, J.~X.} \& \bibinfo{author}{{Wolfe}, A.~M.}
\newblock \bibinfo{journal}{\bibinfo{title}{{Metallicity Evolution in the Early
  Universe}}}.
\newblock {\emph{\JournalTitle{\apjl}}} \textbf{\bibinfo{volume}{533}},
  \bibinfo{pages}{L5--L8}, \doiprefix\url{10.1086/312594}
  (\bibinfo{year}{2000}).
\newblock \eprint{astro-ph/0002513}.

\bibitem{Srianand00}
\bibinfo{author}{{Srianand}, R.}, \bibinfo{author}{{Petitjean}, P.} \&
  \bibinfo{author}{{Ledoux}, C.}
\newblock \bibinfo{journal}{\bibinfo{title}{{The cosmic microwave background
  radiation temperature at a redshift of 2.34}}}.
\newblock {\emph{\JournalTitle{\nat}}} \textbf{\bibinfo{volume}{408}},
  \bibinfo{pages}{931--935}, \doiprefix\url{10.1038/35050020}
  (\bibinfo{year}{2000}).
\newblock \eprint{astro-ph/0012222}.

\bibitem{Dessauges01}
\bibinfo{author}{{Dessauges-Zavadsky}, M.} \emph{et~al.}
\newblock \bibinfo{journal}{\bibinfo{title}{{UVES observations of a damped
  Lyalpha system at z$_{abs}$ = 4.466 towards the quasar APM BR J0307-4945}}}.
\newblock {\emph{\JournalTitle{\aap}}} \textbf{\bibinfo{volume}{370}},
  \bibinfo{pages}{426--435}, \doiprefix\url{10.1051/0004-6361:20010217}
  (\bibinfo{year}{2001}).
\newblock \eprint{astro-ph/0102230}.

\bibitem{Ellison01}
\bibinfo{author}{{Ellison}, S.~L.}, \bibinfo{author}{{Pettini}, M.},
  \bibinfo{author}{{Steidel}, C.~C.} \& \bibinfo{author}{{Shapley}, A.~E.}
\newblock \bibinfo{journal}{\bibinfo{title}{{An Imaging and Spectroscopic Study
  of the z$_{abs}$=3.38639 Damped Ly{\ensuremath{\alpha}} System in Q0201+1120:
  Clues to Star Formation at High Redshift}}}.
\newblock {\emph{\JournalTitle{\apj}}} \textbf{\bibinfo{volume}{549}},
  \bibinfo{pages}{770--779}, \doiprefix\url{10.1086/319457}
  (\bibinfo{year}{2001}).
\newblock \eprint{astro-ph/0010427}.

\bibitem{Molaro01}
\bibinfo{author}{{Molaro}, P.}, \bibinfo{author}{{Levshakov}, S.~A.},
  \bibinfo{author}{{D'Odorico}, S.}, \bibinfo{author}{{Bonifacio}, P.} \&
  \bibinfo{author}{{Centuri{\'o}n}, M.}
\newblock \bibinfo{journal}{\bibinfo{title}{{UVES Observations of QSO
  0000-2620: Argon and Phosphorus Abundances in the Dust-free Damped
  Ly{\ensuremath{\alpha}} System at z$_{abs}$=3.3901}}}.
\newblock {\emph{\JournalTitle{\apj}}} \textbf{\bibinfo{volume}{549}},
  \bibinfo{pages}{90--99}, \doiprefix\url{10.1086/319072}
  (\bibinfo{year}{2001}).
\newblock \eprint{astro-ph/0010434}.

\bibitem{Prochaska01a}
\bibinfo{author}{{Prochaska}, J.~X.}, \bibinfo{author}{{Gawiser}, E.} \&
  \bibinfo{author}{{Wolfe}, A.~M.}
\newblock \bibinfo{journal}{\bibinfo{title}{{Galactic Chemical Abundances at
  z>3. I. First Results from the Echellette Spectrograph and Imager}}}.
\newblock {\emph{\JournalTitle{\apj}}} \textbf{\bibinfo{volume}{552}},
  \bibinfo{pages}{99--105}, \doiprefix\url{10.1086/320467}
  (\bibinfo{year}{2001}).
\newblock \eprint{astro-ph/0101029}.

\bibitem{Prochaska01b}
\bibinfo{author}{{Prochaska}, J.~X.} \emph{et~al.}
\newblock \bibinfo{journal}{\bibinfo{title}{{The UCSD HIRES/Keck I Damped
  Ly{\ensuremath{\alpha}} Abundance Database. I. The Data}}}.
\newblock {\emph{\JournalTitle{\apjs}}} \textbf{\bibinfo{volume}{137}},
  \bibinfo{pages}{21--73}, \doiprefix\url{10.1086/322542}
  (\bibinfo{year}{2001}).
\newblock \eprint{astro-ph/0110350}.

\bibitem{Ledoux02}
\bibinfo{author}{{Ledoux}, C.}, \bibinfo{author}{{Srianand}, R.} \&
  \bibinfo{author}{{Petitjean}, P.}
\newblock \bibinfo{journal}{\bibinfo{title}{{Detection of molecular hydrogen in
  a near Solar-metallicity damped Lyman-alpha system at $z_{abs} \approx 2$
  toward Q 0551-366}}}.
\newblock {\emph{\JournalTitle{\aap}}} \textbf{\bibinfo{volume}{392}},
  \bibinfo{pages}{781--789}, \doiprefix\url{10.1051/0004-6361:20021187}
  (\bibinfo{year}{2002}).

\bibitem{Levshakov02}
\bibinfo{author}{{Levshakov}, S.~A.}, \bibinfo{author}{{Dessauges-Zavadsky},
  M.}, \bibinfo{author}{{D'Odorico}, S.} \& \bibinfo{author}{{Molaro}, P.}
\newblock \bibinfo{journal}{\bibinfo{title}{{Molecular Hydrogen, Deuterium, and
  Metal Abundances in the Damped Ly{\ensuremath{\alpha}} System at
  z$_{abs}$=3.025 toward Q0347-3819}}}.
\newblock {\emph{\JournalTitle{\apj}}} \textbf{\bibinfo{volume}{565}},
  \bibinfo{pages}{696--719}, \doiprefix\url{10.1086/324722}
  (\bibinfo{year}{2002}).
\newblock \eprint{astro-ph/0105529}.

\bibitem{Lopez02}
\bibinfo{author}{{Lopez}, S.}, \bibinfo{author}{{Reimers}, D.},
  \bibinfo{author}{{D'Odorico}, S.} \& \bibinfo{author}{{Prochaska}, J.~X.}
\newblock \bibinfo{journal}{\bibinfo{title}{{Metal abundances and ionization
  conditions in a possibly dust-free damped Lyalpha system at z=2.3}}}.
\newblock {\emph{\JournalTitle{\aap}}} \textbf{\bibinfo{volume}{385}},
  \bibinfo{pages}{778--792}, \doiprefix\url{10.1051/0004-6361:20020181}
  (\bibinfo{year}{2002}).
\newblock \eprint{astro-ph/0202036}.

\bibitem{Petitjean02}
\bibinfo{author}{{Petitjean}, P.}, \bibinfo{author}{{Srianand}, R.} \&
  \bibinfo{author}{{Ledoux}, C.}
\newblock \bibinfo{journal}{\bibinfo{title}{{Molecular hydrogen at $z_{abs}$
  =1.973 toward Q0013-004: dust depletion pattern in damped Lyman
  {\ensuremath{\alpha}} systems}}}.
\newblock {\emph{\JournalTitle{\mnras}}} \textbf{\bibinfo{volume}{332}},
  \bibinfo{pages}{383--391}, \doiprefix\url{10.1046/j.1365-8711.2002.05331.x}
  (\bibinfo{year}{2002}).
\newblock \eprint{astro-ph/0201477}.

\bibitem{Prochaska02}
\bibinfo{author}{{Prochaska}, J.~X.} \emph{et~al.}
\newblock \bibinfo{journal}{\bibinfo{title}{{The UCSD HIRES/Keck I Damped
  Ly{\ensuremath{\alpha}} Abundance Database. IV. Probing Galactic Enrichment
  Histories with Nitrogen}}}.
\newblock {\emph{\JournalTitle{\pasp}}} \textbf{\bibinfo{volume}{114}},
  \bibinfo{pages}{933--954}, \doiprefix\url{10.1086/342354}
  (\bibinfo{year}{2002}).
\newblock \eprint{astro-ph/0206296}.

\bibitem{Centurion03}
\bibinfo{author}{{Centuri{\'o}n}, M.} \emph{et~al.}
\newblock \bibinfo{journal}{\bibinfo{title}{{Early stages of nitrogen
  enrichment in galaxies: Clues from measurements in damped Lyman alpha
  systems}}}.
\newblock {\emph{\JournalTitle{\aap}}} \textbf{\bibinfo{volume}{403}},
  \bibinfo{pages}{55--72}, \doiprefix\url{10.1051/0004-6361:20030273}
  (\bibinfo{year}{2003}).
\newblock \eprint{astro-ph/0302032}.

\bibitem{Ledoux03}
\bibinfo{author}{{Ledoux}, C.}, \bibinfo{author}{{Petitjean}, P.} \&
  \bibinfo{author}{{Srianand}, R.}
\newblock \bibinfo{journal}{\bibinfo{title}{{The Very Large Telescope
  Ultraviolet and Visible Echelle Spectrograph survey for molecular hydrogen in
  high-redshift damped Lyman {\ensuremath{\alpha}} systems}}}.
\newblock {\emph{\JournalTitle{\mnras}}} \textbf{\bibinfo{volume}{346}},
  \bibinfo{pages}{209--228}, \doiprefix\url{10.1046/j.1365-2966.2003.07082.x}
  (\bibinfo{year}{2003}).
\newblock \eprint{astro-ph/0302582}.

\bibitem{Lopez03}
\bibinfo{author}{{Lopez}, S.} \& \bibinfo{author}{{Ellison}, S.~L.}
\newblock \bibinfo{journal}{\bibinfo{title}{{Distinct abundance patterns in
  multiple damped Ly alpha galaxies: Evidence for truncated star formation?}}}
\newblock {\emph{\JournalTitle{\aap}}} \textbf{\bibinfo{volume}{403}},
  \bibinfo{pages}{573--584}, \doiprefix\url{10.1051/0004-6361:20030403}
  (\bibinfo{year}{2003}).
\newblock \eprint{astro-ph/0303441}.

\bibitem{Prochaska03a}
\bibinfo{author}{{Prochaska}, J.~X.}, \bibinfo{author}{{Gawiser}, E.},
  \bibinfo{author}{{Wolfe}, A.~M.}, \bibinfo{author}{{Castro}, S.} \&
  \bibinfo{author}{{Djorgovski}, S.~G.}
\newblock \bibinfo{journal}{\bibinfo{title}{{The Age-Metallicity Relation of
  the Universe in Neutral Gas: The First 100 Damped Ly{\ensuremath{\alpha}}
  Systems}}}.
\newblock {\emph{\JournalTitle{\apjl}}} \textbf{\bibinfo{volume}{595}},
  \bibinfo{pages}{L9--L12}, \doiprefix\url{10.1086/378945}
  (\bibinfo{year}{2003}).
\newblock \eprint{astro-ph/0305314}.

\bibitem{Prochaska03b}
\bibinfo{author}{{Prochaska}, J.~X.}, \bibinfo{author}{{Gawiser}, E.},
  \bibinfo{author}{{Wolfe}, A.~M.}, \bibinfo{author}{{Cooke}, J.} \&
  \bibinfo{author}{{Gelino}, D.}
\newblock \bibinfo{journal}{\bibinfo{title}{{The ESI/Keck II Damped
  Ly{\ensuremath{\alpha}} Abundance Database}}}.
\newblock {\emph{\JournalTitle{\apjs}}} \textbf{\bibinfo{volume}{147}},
  \bibinfo{pages}{227--264}, \doiprefix\url{10.1086/375839}
  (\bibinfo{year}{2003}).
\newblock \eprint{astro-ph/0305312}.

\bibitem{Dessauges04}
\bibinfo{author}{{Dessauges-Zavadsky}, M.}, \bibinfo{author}{{Calura}, F.},
  \bibinfo{author}{{Prochaska}, J.~X.}, \bibinfo{author}{{D'Odorico}, S.} \&
  \bibinfo{author}{{Matteucci}, F.}
\newblock \bibinfo{journal}{\bibinfo{title}{{A comprehensive set of elemental
  abundances in damped Ly{\ensuremath{\alpha}} systems: Revealing the nature of
  these high-redshift galaxies}}}.
\newblock {\emph{\JournalTitle{\aap}}} \textbf{\bibinfo{volume}{416}},
  \bibinfo{pages}{79--110}, \doiprefix\url{10.1051/0004-6361:20034273}
  (\bibinfo{year}{2004}).
\newblock \eprint{astro-ph/0312210}.

\bibitem{Khare04}
\bibinfo{author}{{Khare}, P.} \emph{et~al.}
\newblock \bibinfo{journal}{\bibinfo{title}{{Metals and Dust in
  Intermediate-Redshift Damped Ly{\ensuremath{\alpha}} Galaxies}}}.
\newblock {\emph{\JournalTitle{\apj}}} \textbf{\bibinfo{volume}{616}},
  \bibinfo{pages}{86--109}, \doiprefix\url{10.1086/424893}
  (\bibinfo{year}{2004}).
\newblock \eprint{astro-ph/0408139}.

\bibitem{Turnshek04}
\bibinfo{author}{{Turnshek}, D.~A.} \emph{et~al.}
\newblock \bibinfo{journal}{\bibinfo{title}{{Double-damped
  Ly{\ensuremath{\alpha}} Absorption: A Possible Large Neutral Hydrogen Gas
  Filament near Redshift z=1}}}.
\newblock {\emph{\JournalTitle{\apjl}}} \textbf{\bibinfo{volume}{609}},
  \bibinfo{pages}{L53--L57}, \doiprefix\url{10.1086/422682}
  (\bibinfo{year}{2004}).
\newblock \eprint{astro-ph/0404609}.

\bibitem{Akerman05}
\bibinfo{author}{{Akerman}, C.~J.}, \bibinfo{author}{{Ellison}, S.~L.},
  \bibinfo{author}{{Pettini}, M.} \& \bibinfo{author}{{Steidel}, C.~C.}
\newblock \bibinfo{journal}{\bibinfo{title}{{Zn and Cr abundances in damped
  Lyman alpha systems from the CORALS survey}}}.
\newblock {\emph{\JournalTitle{\aap}}} \textbf{\bibinfo{volume}{440}},
  \bibinfo{pages}{499--509}, \doiprefix\url{10.1051/0004-6361:20052947}
  (\bibinfo{year}{2005}).
\newblock \eprint{astro-ph/0506180}.

\bibitem{Rao05}
\bibinfo{author}{{Rao}, S.~M.}, \bibinfo{author}{{Prochaska}, J.~X.},
  \bibinfo{author}{{Howk}, J.~C.} \& \bibinfo{author}{{Wolfe}, A.~M.}
\newblock \bibinfo{journal}{\bibinfo{title}{{Elemental Abundances in Two High
  Column Density Damped Ly{\ensuremath{\alpha}} Systems at z < 1.5}}}.
\newblock {\emph{\JournalTitle{\aj}}} \textbf{\bibinfo{volume}{129}},
  \bibinfo{pages}{9--19}, \doiprefix\url{10.1086/426365}
  (\bibinfo{year}{2005}).
\newblock \eprint{astro-ph/0409695}.

\bibitem{Dessauges06}
\bibinfo{author}{{Dessauges-Zavadsky}, M.}, \bibinfo{author}{{Prochaska},
  J.~X.}, \bibinfo{author}{{D'Odorico}, S.}, \bibinfo{author}{{Calura}, F.} \&
  \bibinfo{author}{{Matteucci}, F.}
\newblock \bibinfo{journal}{\bibinfo{title}{{A new comprehensive set of
  elemental abundances in DLAs. II. Data analysis and chemical variation
  studies}}}.
\newblock {\emph{\JournalTitle{\aap}}} \textbf{\bibinfo{volume}{445}},
  \bibinfo{pages}{93--113}, \doiprefix\url{10.1051/0004-6361:20053200}
  (\bibinfo{year}{2006}).
\newblock \eprint{astro-ph/0511031}.

\bibitem{Ledoux06}
\bibinfo{author}{{Ledoux}, C.}, \bibinfo{author}{{Petitjean}, P.},
  \bibinfo{author}{{Fynbo}, J.~P.~U.}, \bibinfo{author}{{M{\o}ller}, P.} \&
  \bibinfo{author}{{Srianand}, R.}
\newblock \bibinfo{journal}{\bibinfo{title}{{Velocity-metallicity correlation
  for high-z DLA galaxies: evidence of a mass-metallicity relation?}}}
\newblock {\emph{\JournalTitle{\aap}}} \textbf{\bibinfo{volume}{457}},
  \bibinfo{pages}{71--78}, \doiprefix\url{10.1051/0004-6361:20054242}
  (\bibinfo{year}{2006}).
\newblock \eprint{astro-ph/0606185}.

\bibitem{Meiring06}
\bibinfo{author}{{Meiring}, J.~D.} \emph{et~al.}
\newblock \bibinfo{journal}{\bibinfo{title}{{Elemental abundance measurements
  in low-redshift damped Lyman {\ensuremath{\alpha}} absorbers}}}.
\newblock {\emph{\JournalTitle{\mnras}}} \textbf{\bibinfo{volume}{370}},
  \bibinfo{pages}{43--62}, \doiprefix\url{10.1111/j.1365-2966.2006.10500.x}
  (\bibinfo{year}{2006}).
\newblock \eprint{astro-ph/0604617}.

\bibitem{Dessauges07}
\bibinfo{author}{{Dessauges-Zavadsky}, M.}, \bibinfo{author}{{Calura}, F.},
  \bibinfo{author}{{Prochaska}, J.~X.}, \bibinfo{author}{{D'Odorico}, S.} \&
  \bibinfo{author}{{Matteucci}, F.}
\newblock \bibinfo{journal}{\bibinfo{title}{{A new comprehensive set of
  elemental abundances in DLAs. III. Star formation histories}}}.
\newblock {\emph{\JournalTitle{\aap}}} \textbf{\bibinfo{volume}{470}},
  \bibinfo{pages}{431--448}, \doiprefix\url{10.1051/0004-6361:20077050}
  (\bibinfo{year}{2007}).
\newblock \eprint{0705.1650}.

\bibitem{Ellison07}
\bibinfo{author}{{Ellison}, S.~L.}, \bibinfo{author}{{Hennawi}, J.~F.},
  \bibinfo{author}{{Martin}, C.~L.} \& \bibinfo{author}{{Sommer-Larsen}, J.}
\newblock \bibinfo{journal}{\bibinfo{title}{{Coincident, 100 kpc scale damped
  Ly{\ensuremath{\alpha}} absorption towards a binary QSO: how large are
  galaxies at z \raisebox{-0.5ex}\textasciitilde 3?}}}
\newblock {\emph{\JournalTitle{\mnras}}} \textbf{\bibinfo{volume}{378}},
  \bibinfo{pages}{801--818}, \doiprefix\url{10.1111/j.1365-2966.2007.11847.x}
  (\bibinfo{year}{2007}).
\newblock \eprint{0704.1816}.

\bibitem{Meiring07}
\bibinfo{author}{{Meiring}, J.~D.} \emph{et~al.}
\newblock \bibinfo{journal}{\bibinfo{title}{{New abundance determinations in z
  < 1.5 QSO absorbers: seven sub-DLAs and one DLA}}}.
\newblock {\emph{\JournalTitle{\mnras}}} \textbf{\bibinfo{volume}{376}},
  \bibinfo{pages}{557--572}, \doiprefix\url{10.1111/j.1365-2966.2007.11521.x}
  (\bibinfo{year}{2007}).
\newblock \eprint{astro-ph/0702297}.

\bibitem{Prochaska07}
\bibinfo{author}{{Prochaska}, J.~X.} \emph{et~al.}
\newblock \bibinfo{journal}{\bibinfo{title}{{The UCSD/Keck Damped
  Ly{\ensuremath{\alpha}} Abundance Database: A Decade of High-Resolution
  Spectroscopy}}}.
\newblock {\emph{\JournalTitle{\apjs}}} \textbf{\bibinfo{volume}{171}},
  \bibinfo{pages}{29--60}, \doiprefix\url{10.1086/513714}
  (\bibinfo{year}{2007}).
\newblock \eprint{astro-ph/0702325}.

\bibitem{Noterdaeme08}
\bibinfo{author}{{Noterdaeme}, P.}, \bibinfo{author}{{Ledoux}, C.},
  \bibinfo{author}{{Petitjean}, P.} \& \bibinfo{author}{{Srianand}, R.}
\newblock \bibinfo{journal}{\bibinfo{title}{{Molecular hydrogen in
  high-redshift damped Lyman-{\ensuremath{\alpha}} systems: the VLT/UVES
  database}}}.
\newblock {\emph{\JournalTitle{\aap}}} \textbf{\bibinfo{volume}{481}},
  \bibinfo{pages}{327--336}, \doiprefix\url{10.1051/0004-6361:20078780}
  (\bibinfo{year}{2008}).
\newblock \eprint{0801.3682}.

\bibitem{Wolfe08}
\bibinfo{author}{{Wolfe}, A.~M.}, \bibinfo{author}{{Prochaska}, J.~X.},
  \bibinfo{author}{{Jorgenson}, R.~A.} \& \bibinfo{author}{{Rafelski}, M.}
\newblock \bibinfo{journal}{\bibinfo{title}{{Bimodality in Damped
  Ly{\ensuremath{\alpha}} Systems}}}.
\newblock {\emph{\JournalTitle{\apj}}} \textbf{\bibinfo{volume}{681}},
  \bibinfo{pages}{881--896}, \doiprefix\url{10.1086/588090}
  (\bibinfo{year}{2008}).
\newblock \eprint{0802.3914}.

\bibitem{Jorgenson10}
\bibinfo{author}{{Jorgenson}, R.~A.}, \bibinfo{author}{{Wolfe}, A.~M.} \&
  \bibinfo{author}{{Prochaska}, J.~X.}
\newblock \bibinfo{journal}{\bibinfo{title}{{Understanding Physical Conditions
  in High-redshift Galaxies Through C I Fine Structure Lines: Data and
  Methodology}}}.
\newblock {\emph{\JournalTitle{\apj}}} \textbf{\bibinfo{volume}{722}},
  \bibinfo{pages}{460--490}, \doiprefix\url{10.1088/0004-637X/722/1/460}
  (\bibinfo{year}{2010}).
\newblock \eprint{1008.4676}.

\bibitem{Meiring11}
\bibinfo{author}{{Meiring}, J.~D.} \emph{et~al.}
\newblock \bibinfo{journal}{\bibinfo{title}{{The First Observations of
  Low-redshift Damped Ly{\ensuremath{\alpha}} Systems with the Cosmic Origins
  Spectrograph}}}.
\newblock {\emph{\JournalTitle{\apj}}} \textbf{\bibinfo{volume}{732}},
  \bibinfo{pages}{35}, \doiprefix\url{10.1088/0004-637X/732/1/35}
  (\bibinfo{year}{2011}).
\newblock \eprint{1102.3927}.

\bibitem{Vladilo11}
\bibinfo{author}{{Vladilo}, G.}, \bibinfo{author}{{Abate}, C.},
  \bibinfo{author}{{Yin}, J.}, \bibinfo{author}{{Cescutti}, G.} \&
  \bibinfo{author}{{Matteucci}, F.}
\newblock \bibinfo{journal}{\bibinfo{title}{{Silicon depletion in damped Ly
  {\ensuremath{\alpha}} systems. The S/Zn method}}}.
\newblock {\emph{\JournalTitle{\aap}}} \textbf{\bibinfo{volume}{530}},
  \bibinfo{pages}{A33}, \doiprefix\url{10.1051/0004-6361/201016330}
  (\bibinfo{year}{2011}).

\bibitem{Dutta14}
\bibinfo{author}{{Dutta}, R.} \emph{et~al.}
\newblock \bibinfo{journal}{\bibinfo{title}{{A study of low-metallicity DLAs at
  high redshift and C II* as a probe of their physical conditions}}}.
\newblock {\emph{\JournalTitle{\mnras}}} \textbf{\bibinfo{volume}{440}},
  \bibinfo{pages}{307--326}, \doiprefix\url{10.1093/mnras/stu260}
  (\bibinfo{year}{2014}).
\newblock \eprint{1402.2975}.

\bibitem{Wang15}
\bibinfo{author}{{Wang}, F.} \emph{et~al.}
\newblock \bibinfo{journal}{\bibinfo{title}{{An Ultra-luminous Quasar at z =
  5.363 with a Ten Billion Solar Mass Black Hole and a Metal-rich DLA at z
  {\ensuremath{\sim}} 5}}}.
\newblock {\emph{\JournalTitle{\apjl}}} \textbf{\bibinfo{volume}{807}},
  \bibinfo{pages}{L9}, \doiprefix\url{10.1088/2041-8205/807/1/L9}
  (\bibinfo{year}{2015}).
\newblock \eprint{1506.01765}.

\bibitem{Cooke17}
\bibinfo{author}{{Cooke}, R.~J.}, \bibinfo{author}{{Pettini}, M.} \&
  \bibinfo{author}{{Steidel}, C.~C.}
\newblock \bibinfo{journal}{\bibinfo{title}{{Discovery of the most metal-poor
  damped Lyman-{\ensuremath{\alpha}} system}}}.
\newblock {\emph{\JournalTitle{\mnras}}} \textbf{\bibinfo{volume}{467}},
  \bibinfo{pages}{802--811}, \doiprefix\url{10.1093/mnras/stx037}
  (\bibinfo{year}{2017}).
\newblock \eprint{1701.03103}.

\bibitem{Banados19}
\bibinfo{author}{{Ba{\~n}ados}, E.} \emph{et~al.}
\newblock \bibinfo{journal}{\bibinfo{title}{{A Metal-poor Damped
  Ly{\ensuremath{\alpha}} System at Redshift 6.4}}}.
\newblock {\emph{\JournalTitle{\apj}}} \textbf{\bibinfo{volume}{885}},
  \bibinfo{pages}{59}, \doiprefix\url{10.3847/1538-4357/ab4129}
  (\bibinfo{year}{2019}).
\newblock \eprint{1903.06186}.

\end{thebibliography}
\end{document}